\documentclass[lettersize,journal]{IEEEtran}
\usepackage{amsmath,amsfonts}
\usepackage{algorithmic}
\usepackage{array}
\usepackage[caption=false,font=footnotesize,labelfont=rm,textfont=rm]{subfig}
\usepackage{textcomp}
\usepackage{stfloats}
\usepackage{url}
\usepackage{verbatim}
\usepackage{graphicx}

\usepackage{cite}

\usepackage{ stmaryrd }
\usepackage{amsmath}
\usepackage{amsthm}
\usepackage[flushleft]{threeparttablex}
\usepackage{tabu}
\usepackage{tabularx}
\usepackage{amssymb}

\usepackage{booktabs}
\usepackage{enumerate}
\usepackage{multirow}
\usepackage{xcolor}
\usepackage[dvipsnames, svgnames, x11names]{xcolor}
\usepackage{framed}

\usepackage[linesnumbered,ruled,vlined]{algorithm2e}

\newcommand{\tool}{\texttt{Panther}\xspace}
\newcommand{\toolnott}{Panther\xspace}

\newcommand{\revise}[1]{\textcolor{black}{#1}}

\definecolor{revise-color}{RGB}{0,0,0}

\usepackage{enumitem}

\usepackage{tikz}

\newtheorem{theorem}{Theorem}
\newtheorem{definition}{Definition}

\begin{document}

\title{Panther: A Cost-Effective Privacy-Preserving Framework for GNN Training and Inference Services in Cloud Environments}

\author{Congcong~Chen, Xinyu~Liu, Kaifeng~Huang, Lifei~Wei, and Yang~Shi

\thanks{This article has been accepted for publication in IEEE Transactions on Services Computing. 
\newline \newline
© 2025 IEEE. Personal use of this material is permitted. Permission from IEEE must be obtained for all other uses, in any current or future media, including reprinting/republishing this material for advertising or promotional purposes, creating new collective works, for resale or redistribution to servers or lists, or reuse of any copyrighted component of this work in other works.}
\thanks{Congcong Chen, Xinyu Liu, Kaifeng Huang, and Yang Shi are with the School of Computer Science and Technology, Tongji University, Shanghai 201804, China (e-mail: chencongcong@tongji.edu.cn; 2333088@tongji.edu.cn; kaifengh@tongji.edu.cn; shiyang@tongji.edu.cn).}
\thanks{Lifei Wei is with the College of Information Engineering, Shanghai Maritime University, Shanghai 201306, China (e-mail: lfwei@shmtu.edu.cn).}
\thanks{Corresponding author: Yang~Shi.}
}



\markboth{Journal of \LaTeX\ Class Files,~Vol.~xx, No.~x, October~2025}%
{Chen \MakeLowercase{\textit{et al.}}: \toolnott: A Cost-Effective Privacy-Preserving Framework for GNN Training and Inference Services in Cloud Environments}


 \maketitle

\begin{abstract}

Graph Neural Networks (GNNs) have marked significant impact in traffic state prediction, social recommendation, knowledge-aware question answering and so on. As more and more users move towards cloud computing, it has become a critical issue to unleash the power of GNNs while protecting the privacy in cloud environments. Specifically, the training data and inference data for GNNs need to be protected from being stolen by external adversaries. Meanwhile, the financial cost of cloud computing is another primary concern for users. Therefore, although existing studies have proposed privacy-preserving techniques for GNNs in cloud environments, their additional computational and communication overhead remain relatively high, causing high financial costs that limit their widespread adoption among users.

To protect GNN privacy while lowering the additional financial costs, we introduce \tool, a cost-effective privacy-preserving framework for GNN training and inference services in cloud environments. Technically, \tool leverages four-party computation to asynchronously executing the secure array access protocol, and randomly pads the neighbor information of GNN nodes. We prove that \tool can protect privacy for both training and inference of GNN models. Our evaluation shows that \tool reduces the training and inference time by an average of 75.28\% and 82.80\%, respectively, and communication overhead by an average of 52.61\% and 50.26\% compared with the state-of-the-art, which is estimated to save an average of 55.05\% and 59.00\% in financial costs (based on on-demand pricing model) for the GNN training and inference process on Google Cloud Platform.

\end{abstract}

\begin{IEEEkeywords}
Graph neural networks, GNN model training service, GNN model inference service, privacy-preserving, cloud environments.
\end{IEEEkeywords}

\section{Introduction}
\IEEEPARstart{G}{RAPH} Neural Networks (GNNs) play a crucial role in traffic state prediction~\cite{fang2021spatial}, social recommendation~\cite{song2021social
}, knowledge-aware question answering~\cite{feng2020scalable}, etc. Due to the complexity inherent in machine learning, training and deploying GNN models involve processing large volumes of data and require extensive computational resources. Therefore, it is natural for users to opt for cloud computing as it offers powerful computing capabilities with high scalability, flexibility, and cost-effectiveness. In fact, it has become a popular trend for users to offload computationally intensive tasks in GNN training and inference services to the cloud~\cite{google2024ai,aws2024ai,azure2024cloud}, commonly referred to as Model-as-a-Service (MaaS).

However, data privacy of using GNNs should be managed in cloud environments~\cite{laura2024SAP}. 
For example, social recommendation systems leverage personal information, including personal social circles and interests, to improve user modeling~\cite{song2019session,song2021social,chen2021efficient}. It would cause user privacy leakage, legal issues, and financial losses~\cite{he2021stealing} if the cloud environment is compromised~\cite{Brian2021Another}. Therefore, effective privacy-preserving mechanisms are essential to ensure the secure deployment of GNNs training and inference services in outsourced or cloud-based settings.

\textbf{Background.} To address these concerns, several privacy-preserving GNN frameworks have been proposed, which can be broadly categorized into three paradigms: Federated Learning (FL), Differential Privacy (DP), and cryptography-based solutions. FL-based GNNs~\cite{chen2022vertically,pei2021decentralized} distribute model training across multiple clients without directly sharing raw data. However, recent studies~\cite{wang2019beyond,zhu2019deep} have shown that model updates may still leak sensitive information. DP-based approaches~\cite{daigavane2022node,mueller2022differentially,pei2023privacy-lga-pgnn,zhu2023blink} inject noise to protect data, but this often results in a trade-off between utility and privacy. In contrast, cryptography-based solutions built on Homomorphic Encryption (HE) or Secure Multi-Party Computation (MPC) provide strong formal privacy guarantees by performing computations directly on encrypted or secret-shared data. Among these, HE-based frameworks~\cite{ran2022cryptogcn, peng2024lingcn} for privacy-preserving GNNs often struggle with protecting graph structural information and have high computational costs. In comparison, MPC-based frameworks~\cite{wang2023secgnn,Yuan2024PS-GNN} demonstrates greater practical potential by efficiently handling complex operations without exposing sensitive graph data (including feature data and structural information). Despite their advantages, the significant communication overhead of MPC-based methods remains a key obstacle to their scalable and efficient real-world deployment. 

\revise{Moreover, the financial burden of cloud computing further complicates the adoption of such frameworks. According to CloudZero’s survey~\cite{cloudzero2024cost}, fewer than half of companies in 2024 consider their cloud costs manageable, while 58\% of participants believe their expenses are excessive. As a result, a higher additional cost could discourage user adoption of privacy-preserving GNN frameworks. This inherent trade-off between strong privacy guarantees and system cost-efficiency highlights the need for a new design goal, which we refer to as a \textit{cost-effective privacy-preserving GNN}—a solution that minimizes the accompanying costs while benefiting from privacy-preserving GNN training and inference services.}

\textbf{Challenges.} Typically, a privacy-preserving GNN process involves the computation of the graph (i.e., \textit{neighbor states aggregation}) and representation of the graph (i.e., \textit{adjacency matrix}). Existing solutions suffer from two limitations: First, existing approaches using three-party frameworks are \textit{synchronous} during secure neighbor states aggregation~\cite{wang2023secgnn,Yuan2024PS-GNN}, which means that the next party would wait for the previous party to finish relevant computation. Inevitably, it would bring substantial waiting time, especially in large graphs that require multiple aggregation operations. Second, the state-of-the-art (SOTA)~\cite{wang2023secgnn} uniformly set the maximum degree of nodes in the graph as the lengths of the array of each node to reduce the size of adjacency matrix, bringing extra padding for nodes with fewer neighboring nodes. However, the degrees of nodes in the graph typically exhibit a long-tail distribution, with the majority of nodes having relatively small degrees. For instance, in the CiteSeer dataset, 97.57\% of nodes have degrees less than 10 (see TABLE~XI, Appendix~A), yet the maximum degree reaches 99, making uniform padding largely wasteful.

\textbf{Our solution.} To address these challenges, we propose \tool, a \textit{cost-effective privacy-preserving framework for GNN training and inference services} in cloud environments. Our framework introduces two key innovations: (1) We design a novel four-party protocol that enables \textit{asynchronous} execution of secure neighbor states aggregation. Unlike existing three-party synchronous architectures, our design allows pairwise parallel communication between parties, bypassing the barrier of synchronous neighbor states aggregation; (2) We create an \textit{adjacency matrix} representation of \textit{random neighbor information padding}, which reduces the number of adjacency matrix computations by $\frac{1}{2} \cdot (N \cdot d_{max} + 2E)$ compared to SOTA~\cite{wang2023secgnn}, where $d_{max}$ is the maximum degree, $N$ is the number of nodes, and $E$ is the number of edges. Importantly, our approach maintains full compatibility with standard message-passing GNN models.

\textbf{Evaluation.} We conducted experiments to evaluate the cost-effectiveness of \tool. In terms of efficiency, our experiments show that \tool outperforms SecGNN~\cite{wang2023secgnn}. Compared to SecGNN, the GNN training time using \tool reduced by an average of 75.28\%, while communication overhead decreased by an average of 52.61\%. The GNN inference time using \tool reduced by an average of 82.80\%, while communication overhead decreased by an average of 50.26\%. In terms of financial costs, although the introducing a fourth party increases extra instance leasing fees, the total financial cost (according to \textit{Plan A} in Section~\ref{sec-cost-effectiveness-evaluation}) for GNN training and inference reduced by 55.05\% and 59.00\%, respectively. 
Furthermore, we also compare the efficiency and communication overhead for executing the \textit{secure array access protocol} and \textit{neighbor information padding} standalone, where \tool outperforms SecGNN and and PS-GNN~\cite{Yuan2024PS-GNN} regarding computation time and communication overhead.

In summary, our contributions are as follows:
\begin{itemize}[leftmargin=*]
    \item We proposed a \textit{random neighbor information padding} method to reduce the computational and communication overhead of subsequent protocols while maintaining privacy for GNN neighbor nodes. 
    \item We introduce a fourth party into the protocol to facilitate asynchronous computation in the \textit{secure array access protocol}, thereby eliminating additional waiting time. 
    \item We conducted extensive experiments on three widely-used datasets to evaluate the cost-effectiveness of \tool. 
\end{itemize}
\section{Preliminaries}
This section establishes the necessary background for our work. We start by presenting the key notations in TABLE~\ref{table-notations}, followed by a review of essential concepts in cloud computing, GNNs, and secret sharing techniques.

\begin{table}[!ht]
\color{revise-color}
\caption{Notation Table.}
\resizebox{\linewidth}{!}{
\begin{tabular}{ll}
\toprule
\textbf{Notation} & \textbf{Description} \\
\midrule
\multirow[t]{2}{*}{$\mathcal{G}, N, E, v_i$} & Graph $\mathcal{G}=(\mathcal{V}, \mathcal{E})$ with $N=|\mathcal{V}|$ nodes and $E=|\mathcal{E}|$  \\
     & edges; $v_i \in \mathcal{V}$ is a node    \\
$\mathbf{A}, \mathbf{D}$ & Adjacency matrix; Degree matrix where $\mathbf{D}_{i,i} = \sum_j \mathbf{A}_{i,j}$ \\
$\tilde{\mathbf{A}} = \mathbf{A} + \mathbf{I}$ & Adjacency matrix with self-loops \\
$\tilde{\mathbf{D}}$ & Degree matrix of $\tilde{\mathbf{A}}$ \\
$ne_{v_i,v_j}$ & Neighbor node $v_j$ of node $v_i$ \\
$d_{v_i}, d_{max}$ & Degree of node $v_i$; $d_{max}$ is the maximum degree \\
$D_{in}, D_{out}$ & Respective dimensions of the input and output features \\
$\mathbf{X} \in \mathbb{R}^{N\times D_{in}}$ & Input feature matrix \\
$\mathbf{Z} \in \mathbb{R}^{N\times D_{out}}$ & Output matrix from the GNN model \\
$\mathcal{M} = f(\mathbf{X}, \mathbf{A})$ & GNN model with input $\mathbf{X}$ and adjacency $\mathbf{A}$ \\
$\mathbf{W}^{(l)}$ & Weight matrix at the $l$-th layer of GNN \\
$\sigma$ & Activation function (e.g., ReLU) \\
$sw_{v_i}$ & Sum of edge weights for node $v_i$ \\
$ew_{i,j}$ & Edge weight between node $v_i$ and $v_j$ \\
$\langle x \rangle^m, \llbracket x \rrbracket^m$ & Additive and replicated secret sharing of $x$ modulo $m$ \\
$P_0, P_1, P_2, P_3$ & Four parties involved in secure computation \\
$seed_{01}, seed_{23}$ & Shared randomness between $P_0$/$P_1$ and $P_2$/$P_3$ \\
$\ell$, $p$ & Bit-length of the underlying ring $\mathbb{Z}_{2^\ell}$ and prime $p$ \\
$\ell_x$ & Bit precision for representing value $x$ \\
$\ell_t$ & Fractional precision (used in fixed-point encoding) \\
$r_{01},r^\prime_{01},r_{23},r^\prime_{23}$ & Random value used in secret sharing \\
$a,b,c$ & Constants \\
$\mathbf{x}, \mathbf{y}, \mathbf{z}$ & Column vectors \\ 
$\mathbf{Ne}_{v_i}$, $\mathbf{Ew}_{v_i}$ & Neighbor ID and edge weight arrays for node $v_i$ \\
$\odot$ & Hadamard (element-wise) product \\
$<\cdot,\cdot>$ & Inner product operator \\
$r,t$ & Hyperparameters for inverse square root and Softmax \\
\multirow[t]{2}{*}{$\overline{\sf{trc}}(\cdot,k_1,k_2)$} & Truncates the first $k_2$ bits and the last $k_1$ bits of the input, \\
           & and the results are elements in $\mathbb{Z}_{2^{\ell-k_1-k_2}}$ \\
\bottomrule
\end{tabular}
}
\label{table-notations}
\end{table}

\subsection{Concerns in Cloud Computing}
\textbf{Financial cost concerns.} Cloud computing has become central to enterprise operations, with the Flexera 2024 State of the Cloud Report~\cite{tanner2024cloud} indicating that 71\% of businesses are now \textit{heavy} public cloud users. This trend is prominent even among small to medium-sized businesses, which host 61\% of their workloads and 60\% of their data on public platforms. However, the widespread adoption of cloud services brings the challenge of managing cloud financial costs. The report highlights that 29\% of respondents spend over \$12 million annually on cloud services, making financial cost management a critical concern for enterprises. 

The financial costs for compute instances, storage, and data transfer services are summarized in TABLE~\ref{table-cloud-pricing}. We analyzed the pricing of virtual machines with 4 vCPUs and 16 GB of RAM offered by major cloud providers in Southeast Asia (i.e., Singapore). For GCP, we selected the \texttt{n4-standard-4} machine type with local SSD storage. For Azure, the \texttt{D4s v4} machine type with standard SSD (E10) storage was used, with the redundancy strategy set to LRS. For AWS, the \texttt{m4.xlarge} machine type with general purpose SSD (gp3) storage was used. To simplify the calculation of network traffic, this paper considers only data transfers within the same cloud provider between regions in Asia. In practice, the pricing for data transfer between different cloud providers is usage-based and generally higher than the transfer financial costs within the same provider in the same region.

\begin{table}[!ht]
\resizebox{\linewidth}{!}{
\begin{threeparttable}
\centering
\caption{Financial Cost Comparison for Cloud Computing Services.}
\label{table-cloud-pricing}
\renewcommand{\arraystretch}{1.2}
\begin{tabular}{ccccc}
\toprule
\multicolumn{1}{c}{\multirow{2}{*}{\textbf{Provider}}} & 
\multicolumn{1}{c}{\multirow{2}{*}{\textbf{Instance Type}}} 
& \multicolumn{3}{c}{\textbf{Financial Cost (Plan A/B)$^\ast$}}                                  \\ \cline{3-5} 
\multicolumn{1}{l}{}                          & \multicolumn{1}{l}{}                               & \textbf{Instance} & \textbf{Disk Storage} & \textbf{Data Transfer$^\dagger$} \\
    \midrule
    GCP\cite{google2024compute} & \texttt{n4-standard-4} & \begin{tabular}[c]{@{}c@{}}\$0.2338/\\ \$0.1473\end{tabular} & \begin{tabular}[c]{@{}c@{}}\$0.00012/\\ \$0.00008\end{tabular} & \begin{tabular}[c]{@{}c@{}}\$0.08/\\ \$0.08\end{tabular} \\ \hline
    Azure\cite{azure2024pricing} & \texttt{D4s v4} & \begin{tabular}[c]{@{}c@{}}\$0.192/\\ \$0.134\end{tabular} & \begin{tabular}[c]{@{}c@{}}\$0.01315/\\ \$0.01315\end{tabular} & \begin{tabular}[c]{@{}c@{}}\$0.08/\\ \$0.08\end{tabular} \\ \hline
    AWS\cite{aws2024ec2} & \texttt{m4.xlarge} & \begin{tabular}[c]{@{}c@{}}\$0.25/\\ \$0.1787\end{tabular} & \begin{tabular}[c]{@{}c@{}}\$0.00013/\\ \$0.00013\end{tabular} & \begin{tabular}[c]{@{}c@{}}\$0.09/\\ \$0.09\end{tabular} \\
    \bottomrule
    \end{tabular}%
    \begin{tablenotes} 
        \footnotesize
        \item \textbf{Note:} \textit{Instance Cost} is the cost per hour, \textit{Disk Storage Cost} is the cost per GB per hour, derived by dividing the monthly cost by 730 hours, and \textit{Data Transfer Cost} is the cost per GB for outbound data.
        \item $^\ast$: \textit{Plan A} refers to the on-demand pricing model, while \textit{Plan B} refers to the 1-year resource-based commitment pricing model.
        \item $^\dagger$: Inbound data transfer is free, only outbound data transfer is considered.	
    \end{tablenotes}
\end{threeparttable}
}
\end{table}

\textbf{Security concerns.} Beyond financial cost concerns, security remains a critical issue for cloud users. Reports\cite{tanner2024cloud} indicate that over one-third of respondents plan to transfer all non-sensitive data to the cloud, while nearly one-fifth intend to move all sensitive data to the cloud. However, when sensitive data is compromised, it results in significant financial losses for users. For instance, in recent high-profile data breaches, companies have faced millions of dollars in fines, legal settlements, and reputational damage, underlining the critical importance of safeguarding sensitive cloud data~\cite{shepardson2024att}. Therefore, \textit{preserving sensitive data privacy in cloud environments while reducing financial costs is a pressing challenge}.

\subsection{Graph Neural Networks}
In this paper, we primarily focus on convolutional GNNs, also known as Graph Convolutional Networks (GCNs), which is one of the most commonly used and representative networks in GNNs\cite{kipf2016semi,wang2023secgnn}. A graph $\mathcal{G}=(\mathcal{V},\mathcal{E})$ comprises $N$ nodes $v_i \in \mathcal{V}$ and edges $(v_i,v_j) \in \mathcal{E}$ connecting them. Nodes $v_i$ and $v_j$, connected by the edges $(v_i,v_j)$, are referred to as neighbor nodes. Neighbor nodes of node $v_i \in \mathcal{V}$ are denoted as $ne_{v_i,v_j}$, where $j\in [0,d_{v_i}]$, with $d_{v_i}$ representing the degree of node $v_i$. Graph $\mathcal{G}$ is represented by an adjacency matrix $\mathbf{A} \in \mathbb{R}^{N\times N}$ (binary or weighted) and a degree matrix $\mathbf{D}_{i,i}=\sum_j \mathbf{A}_{i,j}$. The objective of GCNs is to learn a function of features on a graph $\mathcal{G}$. The GCN model $\mathcal{M}=f(\mathbf{X},\mathbf{A})$ takes as inputs a feature matrix $\mathbf{X}$ of size $N\times D_{in}$ and an adjacency matrix $\mathbf{A}$, where $N$ is the number of nodes and $D_{in}$ is the input dimension. The model $\mathcal{M}$ outputs a matrix $\mathbf{Z}$ of size $N\times D_{out}$, where $D_{out}$ is the output dimension.

The forward propagation process of GCN in the $l$-th layer is described as follows:
\begin{align}\label{eq-gcn-forward}
\mathbf{X}^{(l+1)}=f(\mathbf{X}^{(l)},\mathbf{A})=\sigma (\tilde{\mathbf{D}}^{-\frac{1}{2}} \tilde{\mathbf{A}} \tilde{\mathbf{D}}^{-\frac{1}{2}} \mathbf{X}^{(l)} \mathbf{W}^{(l)}),
\end{align} 
where $\mathbf{X}^{(0)}=\mathbf{X}$, $\mathbf{X}^{(L-1)}=\mathbf{Z}$, and $L$ is the number of layers. In Equation ($\ref{eq-gcn-forward}$), $\tilde{\mathbf{A}}=\mathbf{A}+\mathbf{I}$ represents the adjacency matrix $\mathbf{A}$ of the graph $\mathcal{G}$ with self-connections added, where $\mathbf{I}$ is the identity matrix. $\tilde{\mathbf{D}}$ is a diagonal matrix, with $\tilde{\mathbf{D}}_{i,i}=sw_{v_i}=\sum_{j=0}^{N-1} \tilde{\mathbf{A}}_{i,j}=1+\sum_{j=0}^{d_{v_i}-1} ew_{i,j}$, where $ew_{i,j}$ represents the edge weight between node $v_i$ and neighbor node $v_j$, $sw_{v_i}$ represents the sum of edge weights of node $v_i$, and $d_{v_i}$ is the degree of node $v_i$. The activation function $\sigma$, such as $\text{ReLU}(x)=max(0,x)$, is applied in the Equation (\ref{eq-gcn-forward}). $\mathbf{W}^{(l)}$ represents the weight matrix of the GCN model $\mathcal{M}$. 

In this paper, we adopt the same GCN task as in prior works\cite{kipf2016semi,wang2023secgnn}, namely, the semi-supervised multi-classification task. The model consists of two layers, with ReLU serving as the activation function for the hidden layer, and the softmax activation function applied to the last layer to normalize the results, where $\text{Softmax}(x_i)=\frac{e^{x_i}}{\sum_{j=1}^{D_{out}} e^{x_j}}$. Thus, the forward propagation process of the model $\mathcal{M}$ is described as follows: 
\begin{align}\label{eq-gnn-model}
    \mathbf{Z}=f(\mathbf{X},\mathbf{A})=\text{Softmax}(\hat{\mathbf{A}}\text{ReLU}(\hat{\mathbf{A}}\mathbf{X}^{(0)} \mathbf{W}^{(0)})\mathbf{W}^{(1)}),
\end{align}
where $\hat{\mathbf{A}}=\tilde{\mathbf{D}}^{-\frac{1}{2}} \tilde{\mathbf{A}} \tilde{\mathbf{D}}^{-\frac{1}{2}}$ and $\mathbf{X}^{(0)}=\mathbf{X}$. For GCN training, the cross-entropy loss function is employed to compute the error. And the parameters $\mathbf{W}^{(0)}$ and $\mathbf{W}^{(1)}$ are updated through stochastic gradient descent algorithm. 

\subsection{Secret Sharing}\label{sec-secret-sharing}
In this paper, we denote the parties as $P_i$ ($i \in \{0,1,2,3\}$) and design secure protocols based on two distinct secret sharing schemes.

\textbf{2-out-of-2 additive secret sharing.} We denote the 2-out-of-2 additive secret sharing (abbreviated as $(2,2)$-ASS) modulo $m$ as $\langle x \rangle ^m=([x]_0, [x]_1)$, where $m$ is a generic modulus and $x$ is the secret value. Let $r$ be a random value. Then, $[x]_0 = r\pmod m$ and $[x]_1 = x-r\pmod m$, such that the secret value $x$ is expressed as $x \equiv [x]_0 + [x]_1\pmod m$. In our scheme, $P_0$ holds $[x]_0$, $P_1$ holds $[x]_1$, while $P_2$ and $P_3$ do not hold any values. The secret value $x$ can be reconstructed by $P_0$ and $P_1$ exchanging their shares and locally adding them.

\textbf{2-out-of-4 replicated secret sharing.} 
We use $\llbracket x \rrbracket ^m =([x]_0, [x]{_0^\prime}, [x]_1, [x]{_1^\prime}, [x]_2, [x]{_2^\prime}, [x]_3, [x]{_3^\prime})$ denotes 2-out-of-4 replicated secret sharing scheme (abbreviated as $(2,4)$-RSS) modulo $m$. Suppose $P_0$ and $P_1$ share the random seed $seed_{01}$, and $P_2$ and $P_3$ share the random seed $seed_{23}$. To generate a $(2,4)$-RSS, the secret holder first chooses a random value $r$, then generates $(2,2)$-ASS $\langle x \rangle^m$, where $[x]_0=r \pmod m$ and $[x]_1=x-r \pmod m$, then sends $[x]_0$ to $P_0$ and $[x]_1$ to $P_1$. $P_0$ receives $[x]_0$ and calculates $[x]_0^{\prime}=[x]_0-r_{01} \pmod m$, where $r_{01}$ is generated by the random seed $seed_{01}$. Then, $P_0$ sends $[x]_0$ to $P_3$ and $[x]{_0^\prime}$ to $P_2$. $P_1$ receives $[x]_1$ and calculates $[x]{_1^\prime}=[x]_1+r_{01} \pmod m$, and sends $[x]_1$ to $P_2$ and $[x]{_1^\prime}$ to $P_3$. $P_2$ makes $[x]_2=[x]_1, [x]{_2^\prime}=[x]{_0^\prime}$, $P_3$ makes $[x]_3=[x]_0, [x]{_3^\prime}=[x]{_1^\prime}$. 

In the $(2,4)$-RSS, each party $P_i$ holds a pair of secret pieces $([x]_i,[x]{_i^\prime})$, and any two parties $P_i$ and $P_j$, where $i, j\in \{0,1,2,3\}$ and $i \neq j$, can collaboratively reconstruct the secret value $x$. The secret pieces held by each party are detailed in TABLE~\ref{table-secret-holds}. In this paper, the possible values of $m$ include $2^\ell$, $2^{\ell_x}$, $p$, with $2^\ell$ being the default value unless specified otherwise. Here, $\ell$ represents the bit length of the element in $\mathbb{Z}_{2^\ell}$. $\ell_x$ denotes the precision bit length of $x$. Specifically, $x$ is considered positive if $x \in [0, 2^{\ell_x})$, negative if $x \in (2^\ell - 2^{\ell_x},2^\ell) \bmod 2^\ell$, and $p$ is a prime number. We adopt a fixed-point encoding with $t$-bit precision, where $\ell_t$ represents the binary fractional precision of $x$. For simplicity, the modulo operation $(\bmod \ m)$ will be omitted in subsequent sections. Moreover, $\langle x \rangle ^m$ and $\llbracket x \rrbracket ^m$ will be abbreviated as $\langle x \rangle$ and $\llbracket x \rrbracket$, respectively.

\begin{table}[!ht]
\centering
\renewcommand{\arraystretch}{1.2}
\caption{The Secret Pieces Held by the Parties in the $(2,4)$-RSS.}
\label{table-secret-holds}
\begin{tabular}{ccccc}
\toprule
\bfseries{Secret pieces} & \bfseries{$P_0$} & \bfseries{$P_1$} & \bfseries{$P_2$} & \bfseries{$P_3$} \\ 
\midrule
    $[x]_i$ & $r$ & $x-r$ & $x-r$ & $r$ \\
    \hline
    $[x]{_i^\prime}$ & $r-r_{01}$ & $x-r+r_{01}$ & $r-r_{01}$ & $x-r+r_{01}$ \\
\bottomrule
\end{tabular}%
\end{table}

\section{\toolnott}

\subsection{Overview}
\begin{figure}[!h]
    \centering
    \includegraphics[width=\linewidth]{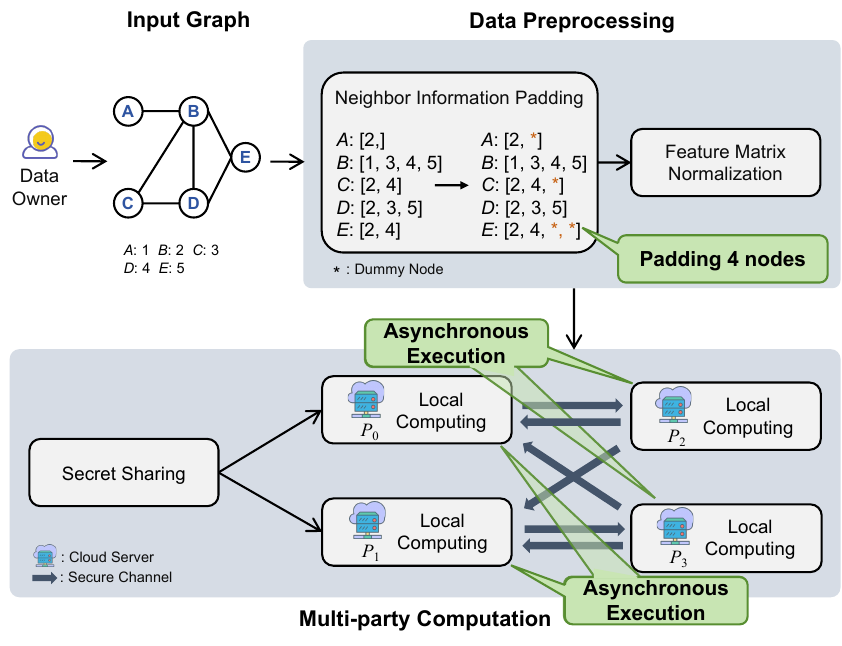}
\caption{GNN training/inference process for \toolnott (using the secure array access protocol as an example in the multi-party computation phase).} 
\label{fig-overview}
\end{figure}

As depicted in Fig.~\ref{fig-overview}, our system consists of two types of entities: \textit{data owners} and \textit{cloud servers}.
\begin{itemize}[leftmargin=*]
    \item \textbf{Data owners.}  Data owners have data relevant to GNN training and inference, which may include online shopping enterprises, social media service providers, and others. They share their private data with cloud servers to train GNN models or for inference purposes. During the training and inference phases of GNN, cloud servers must ensure the confidentiality of data owners' private information, including features, graph structure, 
    and inference results.
    \item \textbf{Cloud servers.} Our secure computation framework consists of four cloud servers (also referred to as four parties in this paper), denoted as $P_0, P_1, P_2$, and $P_3$, involved in the protocol computation. These parties secretly train the GNN model using secure protocols based on secret sharing and utilize the trained model to offer inference services.
\end{itemize}

The overall system design is divided into two phases: \textit{data preprocessing phase}, and \textit{multi-party computation phase}. Data preprocessing phase mainly includes neighbor information padding and feature matrix normalization. The data owner inputs the graph to the cloud servers through the data preprocessing phase. During multi-party computation phase, all cloud servers cooperatively perform privacy-preserving training and inference computation of GNNs based on secure protocols. 
Our system aims to achieve four key privacy goals: (1) protecting the feature matrix $\mathbf{X}$ of the data owner, (2) hiding the adjacency matrix $\mathbf{A}$ (including neighbor IDs, edge weights, and node degrees), (3) keeping the model weights $\mathbf{W}^{(0)}$ and $\mathbf{W}^{(1)}$ confidential from the cloud, and (4) preserving the privacy of the inference outputs $\mathbf{Z}$.

\subsection{System Design}\label{sec-system-design}
\subsubsection{Data preprocessing} 
Unlike Convolutional Neural Networks (CNNs), in GNNs, the model’s input information usually includes a feature matrix $\mathbf{X}$ and an adjacency matrix $\mathbf{A}$. The feature matrix $\mathbf{X}$ contains the private information of nodes, while the adjacency matrix $\mathbf{A}$ contains comprehensive graph structure information, including node neighbor IDs $ne_{v_i,v_j}$, edge weights $ew_{v_i,v_j}$, and node degrees $d_{v_i}$, where $i$ and $j$ denote node indices. Therefore, for training and inference of GNN models, it is necessary to encrypt the adjacency matrix $\mathbf{A}$ before inputting the data into the model. Additionally, to accelerate convergence and improve accuracy, we typically need to normalize the data.

\begin{figure}[!h]
\centering
    \includegraphics[width=\linewidth]{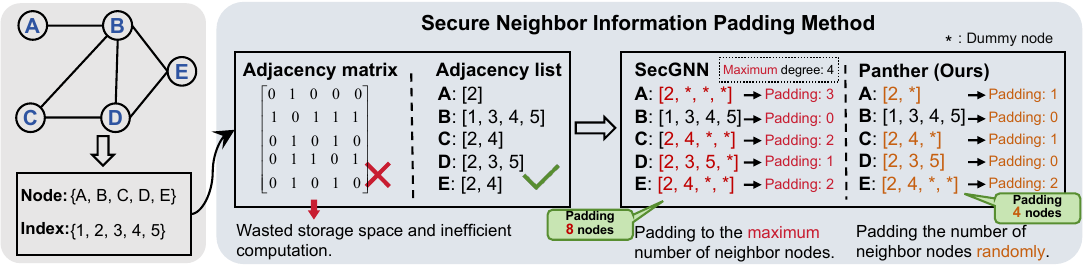}
\caption{\revise{Similarities and differences between the secure neighbor information padding methods of \toolnott and SecGNN.}}
\label{fig-padding-method}
\end{figure}

\textbf{Neighbor information padding}. The adjacency matrix $\mathbf{A}$ in graph $\mathcal{G}$ is typically sparse\cite{geng2020awb,auten2020hardware,hu2020featgraph,huang2020ge,li2021gcnax}, and encrypting the entire matrix would result in considerable communication overhead. To mitigate this challenge, SecGNN\cite{wang2023secgnn} proposes encrypting the adjacency matrix $\mathbf{A}$ in an array-like format, as illustrated in Fig.~\ref{fig-padding-method}. SecGNN prevents the disclosure of graph neighbor information by padding dummy nodes and dummy data (i.e., zeros) in the neighbor ID array $\mathbf{Ne}_{v_i}$ and the edge weight array $\mathbf{Ew}_{v_i}$. They extend all neighbor ID arrays and edge weights to match the length of the maximum degree $d_{max}$, as illustrated in Equation (\ref{eq-secgnn-ne}) and Equation (\ref{eq-secgnn-ew}):
\begin{align}
    \label{eq-secgnn-ne}
    \mathbf{Ne}_{v_i}&=\{ne_{v_i,0},\cdots,ne_{v_i,{d_{v_i}-1}}\} \notag\\
    & \quad \bigcup \{ne_{v_i,0}^\prime,\cdots,ne_{v_i,{d_{max}-d_{v_i}-1}}^\prime\},
\end{align}
\vspace{-0.5cm}
\begin{align}
    \label{eq-secgnn-ew}
    \mathbf{Ew}_{v_i}&=\{ew_{v_i,0},\cdots,ew_{v_i,{d_{v_i}-1}}\} \bigcup 
    \{0,\cdots,0\}.
\end{align} 

However, this method introduces redundant computational and communication overhead. 
For instance, in social networks, some users have many friends, while others have only a few, leading to substantial additional communication overhead. As show in Fig.~\ref{fig-padding-method}, our \textit{random neighbor information padding} method enhances overall performance by reducing the computational and communication overhead for each node by approximately $\frac{1}{2}\cdot (d_{max}-d_{v_i})$. Instead of expanding the neighbor ID array and edge weight array of each node to the size of $d_{max}$, our method extends them to the length of $d_{v_i}^k$, where $k\in [0,d_{max}-d_{v_i}-1]$ is a \textit{random} value. As $k$ is randomly chosen, the expectation for node $v_i$ is $E_{v_i}^k=\frac{1}{2}\cdot (d_{max}-d_{v_i})$. Let $C_1$ denote the total number of additional computations required for SecGNN, given by $C_1=\sum_{i=0}^{N-1} (d_{max}-d_{v_i})$. For our padding method, the total number of additional computations $C_2$ is given by $C_2=\sum_{i=0}^{N-1} E_{{v_i}}^k$, resulting in a reduction of $C=C_1-C_2=\frac{1}{2} \cdot \sum_{i=0}^{N-1} (d_{max}-d_{v_i})$ computations. Since the total degree of the graph is $\sum_{i=0}^{N-1} d_{v_i} = 2E$, where $E$ represents the number of edges in the graph, the reduced computation can be written as $C = \frac{1}{2} \cdot (N \cdot d_{max} + 2E)$. No information about the graph $\mathcal{G}$ will be leaked since $k$ is chosen at random. 
The efficiency and security analysis of our method is shown in Appendix~A.

\textbf{Feature matrix normalization}. For the sake of speeding up the model convergence and improving the accuracy, the features of each node are usually normalized before the training and inference of the GNN model, as Equation (\ref{eq-normalization}), where $D_{in}$ is the feature dimension:
\begin{align}\label{eq-normalization}
    x_{i} = \frac{x_i}{\sum_{j=0}^{D_{in}} x_j}, i\in \{0, \cdots, D_{in}-1\}.
\end{align}

The data preprocessing process can be summarized as follows: Before computation, data owners (1) pad the neighbor ID and edge weight arrays to protect the graph structure, and (2) normalize the feature matrix $\mathbf{X}$. Subsequently, the secret shares of the normalized $\mathbf{X}$, $\mathbf{Ne}_{v_i}$ and $\mathbf{Ew}_{v_i}$ are sent to the cloud servers for further computation. The subsequent computations involving the graph structure in GNNs can be considered as operations on arrays.

\subsubsection{Multi-party computation}
\tool uses secret sharing techniques to protect data privacy and completes the privacy-preserving GNN training and inference processes through secure protocols based on secret sharing, including basic operations and protocols for matrix multiplication, array access, ReLU, Softmax, division, and inverse square root.

\textbf{Secret sharing}. To protect the graph information from being leaked, data owners transmit the node feature matrix $\mathbf{X}$, the neighbor ID array $\mathbf{Ne}_{v_i}$, and the edge weight array $\mathbf{Ew}_{v_i}$ to the cloud servers using secret sharing for secure multi-party computation. In this paper, we employ 2-out-of-2 additive secret sharing and 2-out-of-4 replicated secret sharing schemes to ensure the privacy of the graph, see Section \ref{sec-secret-sharing} for details.

\textbf{Privacy-preserving GNN training and inference}. After secret sharing, all parties perform the privacy-preserving training and inference of the GNN collaboratively based on secure protocols. 
Equation (\ref{eq-gnn-model}) delineates the forward propagation process of the GNN model $\mathcal{M}$. Performing the forward propagation process enables GNN inference. The training process can be decomposed into the following steps:

(1) The aggregate state $\bar{\mathbf{x}}_{v_i}^{(l)}$ for node $v_i$ in the $l$-th layer can be computed using Equation (\ref{eq-gnn-ne-state-l}):
\begin{align}
    \label{eq-gnn-ne-state-l}
    \bar{\mathbf{x}}_{v_i}^{(l)}=(\hat{\mathbf{A}}\mathbf{X}^{(l)})_{v_i}=(\tilde{\mathbf{D}}^{-\frac{1}{2}} \tilde{\mathbf{A}} \tilde{\mathbf{D}}^{-\frac{1}{2}}\mathbf{X}^{(l)})_{v_i},
\end{align}
where $(\cdot)_{v_i}$ represents the vector corresponding to the row of node $v_i$ in the matrix $\hat{\mathbf{A}}\mathbf{X}^{(l)}$. Because $\tilde{\mathbf{D}}$ is a diagonal matrix, $sw_{v_i}=\tilde{\mathbf{D}}_{i,i}$ and $\tilde{\mathbf{A}}_{i,j}=0$ when node $v_j$ is not a neighbor of $v_i$. Thus, Equation~(\ref{eq-gnn-ne-state-l}) simplifies as follows \cite{wang2023secgnn}:
\begin{align}
 \bar{\mathbf{x}}_{v_i}^{(l)}&=(\tilde{\mathbf{D}}^{-\frac{1}{2}} \tilde{\mathbf{A}} \tilde{\mathbf{D}}^{-\frac{1}{2}}\mathbf{X}^{(l)})_{v_i} 
 =\sum_{j=0}^{N-1} \frac{\tilde{\mathbf{A}}_{i,j}}{\sqrt{\tilde{\mathbf{D}}_{i,i}\tilde{\mathbf{D}}_{j,j}}} \mathbf{X}_j^{(l)} 
 \notag \\
 &= \frac{1}{sw_{v_i}}{\mathbf{x}_{v_i}^{(l)}}+\sum_{j=0}^{d_{v_i}^\prime} \frac{\mathbf{Ew}_{v_i,j}}{\sqrt{sw_{v_i}}\sqrt{sw_{\mathbf{Ne}_{v_i,j}}}}\mathbf{x}_{\mathbf{Ne}_{v_i,j}}^{(l)}.
\end{align}
Here, $\mathbf{x}_{v_i}^{(l)}$ represents the state of node $v_i$ at the $l$-th layer, $d_{v_i}^\prime$ denotes the degree after adding dummy neighbor nodes, and $\mathbf{x}_{\mathbf{Ne}_{v_i,j}}^{(l)}$ denotes the state of the $j$-th neighbor node of $v_i\in \mathcal{V}$ at the $l$-th layer.

(2) Compute the $\text{ReLU}(\cdot)$ and $\text{Softmax}(\cdot)$ activation functions, as well as the cross-entropy loss $\mathcal{L}$, as described in Equation~(\ref{eq-gnn-model}).

(3) Update the $l$-th layer weights $\mathbf{W}^{(l)}$ of the GNN model using the stochastic gradient descent algorithm, where $\alpha$ denotes the learning rate and $\mathcal{L}$ is the loss value:
\begin{align}
    \label{eq-gradient-descent-algo}
    \mathbf{W}^{(l)}=\mathbf{W}^{(l)}-\alpha\frac{\partial \mathcal{L}}{\partial \mathbf{W}^{(l)}}.
\end{align}

\subsection{Threat Model}
Our scenario is that data owners outsource their data to cloud servers for GNN model training and inference. 
Similar to prior works\cite{mohassel2017secureml,mohassel2018aby3,wagh2021falcon,wang2023secgnn}, we assume that the protocol works under the semi-honest setting and that the cloud servers do not collude with each other. This means that although the adversary follows the protocol honestly, their goal is to maximize the extraction of private information from data owners. This aligns with real-world scenarios because cloud service providers like GCP, Azure, and AWS are unlikely to deviate from agreements to avoid damaging their reputation. We assume that secure communication channels exist between these four cloud servers, preventing the adversary from learning any information during communication process. Our main goal is to preserve the privacy of the GNN training and inference phases. Adversarial, backdoor, and other related AI attacks on GNN models are beyond the scope of this work.

\subsection{Security Analysis}
We prove the security of our protocol against a semi-honest, non-colluding adversary using the standard real-ideal paradigm in the hybrid model~\cite{canetti2000security,canetti2001universally}. This ensures that the view of any party during the real protocol execution is computationally indistinguishable from a simulation in an ideal world. For brevity, the formal definitions and ideal functionalities are provided in Appendix~B, and detailed security proofs are presented in Appendix~C.
\section{Protocol Constructions}
This section details the specific implementation of the secure protocols involved in the computations discussed in Section~\ref{sec-system-design}. This includes the basic operations and protocols for secure matrix multiplication, secure array access, ReLU, softmax, division, and inverse square root. The theoretical complexity of our protocol is shown in Appendix~D.

\subsection{Basic Operations}
\textbf{Share conversion.} To convert $(2,2)$-ASS to $(2,4)$-RSS, each party $P_i$ ($i\in \{0,1,2,3\}$) executes steps similar to those in the generation of $(2,4)$-RSS. 
For converting $(2,4)$-RSS to $(2,2)$-ASS, parties $P_0$ and $P_1$ need only to add the random values $r_{01}^\prime$ and $-r_{01}^\prime$ to their shares $[x]_0$ and $[x]_1$, respectively. The share conversion protocol is denoted by $\Pi_{\mathsf{SC}}$. 

\textbf{Reconstruction of $\langle x \rangle$ or $\llbracket x \rrbracket$.} The secret value $x$ in $\langle x \rangle$ can be reconstructed by exchanging $P_0$ and $P_1$ and then locally summing them. For $\llbracket x \rrbracket$, any two parties $P_i(i \in \{0,1,2,3\})$ can reconstruct $x$ by exchanging an element and summing it locally. For example, if $P_0$ and $P_2$ wish to reconstruct $x$, they simply exchange $[x]_i$, whereas if $P_0$ and $P_3$ intend to reconstruct $x$, they merely exchange $[x]{_i^\prime}$. We denote the reconstruction protocol as $\Pi_{\mathsf{Rec}}$. 

\textbf{Secure addition and subtraction.} The addition and subtraction operations for $\langle x \rangle$ and $\llbracket x \rrbracket$ are computationally free, allowing each party to compute them locally without communication. In other words, $\langle x \pm y \rangle = \langle x \rangle \pm \langle y \rangle$ and $\llbracket x \pm y \rrbracket = \llbracket x \rrbracket \pm \llbracket y \rrbracket$. After each party locally computes $\langle x \pm y \rangle$ and $\llbracket x \pm y \rrbracket$, the resulting secret pieces also adhere to the constraints of $(2,2)$-ASS and $(2,4)$-RSS. 

\textbf{Linear operations.} Given public constants $a$, $b$, and $c$, the linear operation $\llbracket ax + by + c \rrbracket$ can be computed locally. $P_0$ computes $(a[x]_0 + b[y]_0 + c, a[x]{_0^\prime} + b[y]{_0^\prime} + c)$ locally, and $P_i$ computes $(a[x]_i + b[y]_i, a[x]{_i^\prime} + b[y]{_i^\prime})$ locally, where $i \in \{1,2,3\}$. The process of linear operations for $\langle x \rangle$ is similar to that for $\llbracket x \rrbracket$. Specifically, $P_0$ computes $a \langle x \rangle_0 + b\langle y \rangle_0 + c$, $P_1$ computes $a \langle x \rangle_1 + b\langle y \rangle_1$, while $P_2$ and $P_3$ are inactive. 

\textbf{Truncation.} In this paper, we adopt Bicoptor 2.0's recommendation \cite{zhou2023bicoptor2} of employing distinct truncation protocols for various operations. For linear layer operations like multiplication, we implement the probabilistic truncation protocol proposed by SecureML\cite{mohassel2017secureml} and utilize the \emph{truncate-then-multiply} technique (Algorithm 3 in Bicoptor 2.0) to mitigate truncation failures. For ReLU, we employ the non-interactive deterministic truncation protocol proposed by Bicoptor 2.0 (Algorithm 5) to prevent truncation failures. Implementation details for truncation will be provided later.

\textbf{Secure multiplication.} We implement $\Pi_{\mathsf{Mult}}$ based on the $(2,4)$-RSS secure multiplication protocol proposed by PrivPy~\cite{li2019privpy}, as presented in \textsc{Protocol}~\ref{algo_mult}. This protocol differs from approaches like SecureML\cite{mohassel2017secureml} and ABY3\cite{mohassel2018aby3} by not utilizing the online-offline paradigm. This minimizes the overall communication overhead, eliminating computation or communication overhead during the offline phase. Here, we utilize the \emph{truncate-then-multiply} technique, which involves truncating the secret-shared value before performing the multiplication operation, to prevent truncation failures. Initially, $P_i(i\in \{0,1,2,3\})$ performs local truncation, ensuring that $[x]_i = [x]_i/2^{\ell_t},[x]{_i^\prime} = [x]_i^\prime/2^{\ell_t}, [y]_i = [y]_i/2^{\ell_t},[y]{_i^\prime} = [y]_i^\prime/2^{\ell_t}$, before proceeding with the remaining local computations.  
Protocol $\Pi_{\mathsf{Mult}}$ can be completed in 2 rounds of online communication. 
Appendix~C presents the proof of Theorem~\ref{theorem-mult}.

\begin{theorem}
\label{theorem-mult}
\textit{Protocol $\Pi_{\mathsf{Mult}}$ securely realizes $\mathcal{F}_{\mathsf{Mult}}$ (see Fig.~8, Appendix~B) in the presence of one semi-honest corrupted party}.
\end{theorem}

\begin{algorithm}
\SetKw{return}{return}\SetKwInOut{Entities}{Entities}\SetKwInOut{Input}{Input}\SetKwInOut{Output}{Output}

\caption{Secure Multiplication $\Pi_{\mathsf{Mult}}$}
\label{algo_mult} 
\Input{Secret shared pieces of $\llbracket x \rrbracket$ and $\llbracket y \rrbracket$.} 
\Output{$\llbracket z \rrbracket$, where $z = x/ 2^{\ell_t} \cdot y / 2^{\ell_t}$.}

 $P_i (i \in \{0,1,2,3\})$ performs truncation locally such that $[x]_i = [x]_i/2^{\ell_t}, [x]{_i^\prime} = [x]_i^\prime/2^{\ell_t}, [y]_i = [y]_i/2^{\ell_t}, [y]{_i^\prime} = [y]_i^\prime/2^{\ell_t}$\;

 $P_0,P_1$ generates $r_{01}$ and $r_{01}^\prime$ from $seed_{01}$, $P_2,P_3$ generates $r_{23}$ and $r_{23}^\prime$ from $seed_{23}$.
 
 $P_0$ let $[s]_0=[x]_0[y]{_0^\prime}-r_{01}$, $[s]{_0^\prime}=[x]{_0^\prime}[y]_0-r_{01}^\prime$\;
 
 $P_1$ let $[s]_1=[x]_1[y]{_1^\prime}+r_{01}$, $[s]{_1^\prime}=[x]{_1^\prime}[y]_1+r_{01}^\prime$\;
 
 $P_2$ let $[s]_2=[x]_2[y]{_2^\prime}-r_{23}$, $[s]{_2^\prime}=[x]{_2^\prime}[y]_2-r_{23}^\prime$\;
 
 $P_3$ let $[s]_3=[x]_3[y]{_3^\prime}+r_{23}$, $[s]{_3^\prime}=[x]{_3^\prime}[y]_3+r_{23}^\prime$\;
 $P_0$ sends $[s]_0$ to $P_3$ and $[s]{_0^\prime}$ to $P_2$, $P_1$ sends $[s]_1$ to $P_2$ and $[s]{_1^\prime}$ to $P_3$, $P_2$ sends $[s]_2$ to $P_1$ and $[s]{_2^\prime}$ to $P_0$, $P_3$ sends $[s]_3$ to $P_0$ and $[s]{_3^\prime}$ to $P_1$\;

 $P_0$ sets $[z]_0=([s]_0+[s]_3)$ and $[z]{_0^\prime}=([s]{_0^\prime}+[s]{_2^\prime})$\;
 $P_1$ sets $[z]_1=([s]_1+[s]_2)$ and $[z]{_1^\prime}=([s]{_1^\prime}+[s]{_3^\prime})$\;
 $P_2$ sets $[z]_2=([s]_2+[s]_1)$ and $[z]{_2^\prime}=([s]{_2^\prime}+[s]{_0^\prime})$\;
 $P_3$ sets $[z]_3=([s]_3+[s]_0)$ and $[z]{_3^\prime}=([s]{_3^\prime}+[s]{_1^\prime})$\;
\end{algorithm} 

\subsection{Building Blocks}
\subsubsection{Hadamard product protocol} To compute $\llbracket \mathbf{z} \rrbracket$, where $\mathbf{z}=\mathbf{x}\odot \mathbf{y}$ with $\mathbf{x}$, $\mathbf{y}$, and $\mathbf{z}$ as vectors, and $\odot$ denoting the Hadamard product. We can extend the $\Pi_{\mathsf{Mult}}$ protocol simply by applying it to each element of the matrix, thereby achieving the $\Pi_{\mathsf{HadamardProd}}$ protocol. The $\Pi_{\mathsf{HadamardProd}}$ protocol also involves only 2 rounds of communication. 
Appendix~C provides the proof of Theorem \ref{theorem-hadamard-prod}.

\begin{theorem}
\label{theorem-hadamard-prod}
\textit{Protocol $\Pi_{\mathsf{HadamardProd}}$ securely realizes $\mathcal{F}_{\mathsf{HadamardProd}}$ (see Fig.~9, Appendix~B) in the presence of one semi-honest corrupted party in the ($\mathcal{F}_{\mathsf{Mult}}$)-hybrid model}.
\end{theorem}

\subsubsection{Secure matrix multiplication protocol} The protocol $\Pi_{\mathsf{MatMult}}$ is similar to the protocol $\Pi_{\mathsf{Mult}}$ and achieves $\llbracket \mathbf{Z} \rrbracket = \llbracket \mathbf{X} \cdot \mathbf{Y} \rrbracket$ by expanding the individual elements into matrices, where $\mathbf{X}$, $\mathbf{Y}$, and $\mathbf{Z}$ are matrices. 
The proof of Theorem \ref{theorem-matmult} is provided in Appendix~C.

\begin{theorem}
\label{theorem-matmult}
\textit{Protocol $\Pi_{\mathsf{MatMult}}$ securely realizes $\mathcal{F}_{\mathsf{MatMult}}$ (see Fig.~10, Appendix~B) in the presence of one semi-honest corrupted party in the ($\mathcal{F}_{\mathsf{Mult}}$)-hybrid model}.
\end{theorem}

\subsubsection{Secure array access protocol} 
Aggregating encrypted neighbor states presents a significant challenge. As both neighbor IDs and states are encrypted via secret sharing techniques, the resulting access must remain encrypted. Prior work~\cite{wang2023secgnn} addresses this issue by reformulating neighbor states aggregation as an array access problem. Existing frameworks~\cite{wang2023secgnn,Yuan2024PS-GNN} improve the secure array access protocol of Blanton et al.~\cite{blanton2020improved}, reducing communication to transmitting only $t+1$ elements per party in each round, where $t$ is the array length.

\begin{figure}[!h]
\centering
 \includegraphics[width=\linewidth]{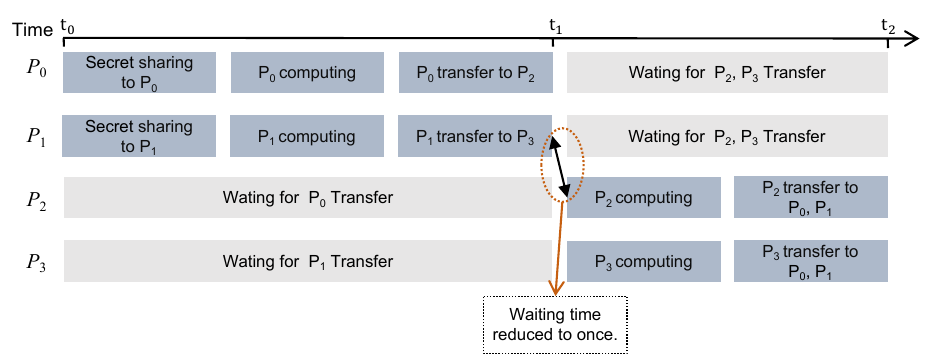}
\caption{Secure array access protocol for \toolnott.} 
\label{fig-aa}
\end{figure}

However, these secure array access protocol operates synchronously. 
The protocol requires cooperation among three parties, where $P_1$ computes after $P_0$, and $P_2$ follows. To enhance secure computation performance, we intuitively eliminate these waiting periods. By incorporating a fourth party, we eliminate the waiting time between $P_1$ and $P_2$, as depicted in Fig.~\ref{fig-aa}. In our protocol $\Pi_{\mathsf{AA}}$, $P_0$ and $P_1$, as well as $P_2$ and $P_3$, execute asynchronously. Our \textit{secure array access protocol} eliminates the extra wait time for single computations, improving the computation time by an average of 36.18\% (see Section~\ref{sec-aa-performence-evaluation} for details).

Assume that an array $\mathbf{a}$ of length $t$ has been encrypted as $\llbracket \mathbf{a} \rrbracket$ using $(2,4)$-RSS, and the index $I$ has been encrypted as $\llbracket I \rrbracket$. Below, we describe the details of $\Pi_{\mathsf{AA}}$:
\begin{enumerate}[leftmargin=*]
    \item \textit{Share conversion} ($\llbracket \cdot \rrbracket$ to $\langle \cdot \rangle$). To enable asynchronous computation for the protocol $\Pi_{\mathsf{AA}}$, we first use $\Pi_{\mathsf{SC}}$ to convert the shares $\llbracket \mathbf{a} \rrbracket$ and $\llbracket I \rrbracket$ into $\langle \mathbf{a} \rangle$ and $\langle I \rangle$.
    \item \textit{Element rotation.} \textbf{a)} $P_0$ and $P_1$ generate random numbers $r_{01}^{a}, \mathbf{r}_{01}^{b}=[r_{01}^{0},\cdots,r_{01}^{t-1}]$ from $seed_{01}$, while $P_2$ and $P_3$ generate random numbers $r_{23}^{a}, r_{23}^{b}$ from $seed_{23}$. Here, $r_{01}^{a}$ all need to be modulo $t$, where $t$ represents the length of the secret array $\llbracket \mathbf{a} \rrbracket$.
    
    \textbf{b)} Party $P_0$ and $P_1$ rotate the location of their secret-shared $[\mathbf{a}]_i$ by $r_{01}^{a}$, where $i \in \{0,1\}$:
    \begin{align}\label{aa-element-rotation}
    [\mathbf{a}]_i^\prime & = [\mathbf{a}]_i \circlearrowright r_{01}^{a}=\{[\mathbf{a}[0]]_i, \cdots ,[\mathbf{a}[t-1]]_i\} \circlearrowright r_{01}^{a} \notag \\
    & = \{[\mathbf{a}[t-1-r_{01}^{a}]]_i, \cdots, [\mathbf{a}[t-1]]_i, [\mathbf{a}[0]]_i, \cdots, \notag \\
    & \qquad [\mathbf{a}[t-r_{01}^{a}]]_i\};
    \end{align}
    The secret index $[I]_i$ held by party $P_i$ is offset by $r_{01}^{a}$ locations, such that the offset indices are $[j]_0 = ([I]_0+r_{01}^{a})$ and $[j]_1=([I]_1+r_{01}^{a})$, where $i \in \{0,1\}$.

    \textbf{c)} To maintain the privacy of the array elements, $P_0$ and $P_1$ must mask each secret-shared array element in $[\mathbf{a}]_i^\prime$ with $\mathbf{r}_{01}^{b}$, which was generated in step a):
    \begin{align}
    [\mathbf{a}]_0^{\prime\prime} & = [\mathbf{a}]_0^\prime \oplus \mathbf{r}_{01}^{b} \notag \\
    & = \{[\mathbf{a}[t-1-r_{01}^{a}]]_0+\mathbf{r}_{01}^{b}[t-1-r_{01}^{a}], \cdots, \notag \\
    &  \qquad [\mathbf{a}[t-1]]_0+\mathbf{r}_{01}^{b}[t-1],[\mathbf{a}[0]]_0+\mathbf{r}_{01}^{b}[0], \cdots, \notag \\
    &  \qquad [\mathbf{a}[t-r_{01}^{a}]]_0+\mathbf{r}_{01}^{b}[t-r_{01}^{a}]\};
    \end{align}
    \vspace{-0.5cm}
    \begin{align}
    [\mathbf{a}]_1^{\prime\prime} & = [\mathbf{a}]_1^\prime \ominus \mathbf{r}_{01}^{b} \notag \\
    & = \{[\mathbf{a}[t-1-r_{01}^{a}]]_1-\mathbf{r}_{01}^{b}[t-1-r_{01}^{a}], \cdots, \notag \\
    &  \qquad [\mathbf{a}[t-1]]_1-\mathbf{r}_{01}^{b}[t-1],[\mathbf{a}[0]]_1-\mathbf{r}_{01}^{b}[0], \cdots, \notag \\
    &  \qquad [\mathbf{a}[t-r_{01}^{a}]]_1-\mathbf{r}_{01}^{b}[t-r_{01}^{a}]\};
    \end{align}
    
    \textbf{d)} $P_0$ sends $[\mathbf{a}]_0^{\prime\prime}$ and $[j]_0$ to $P_2$, and $P_1$ sends $[\mathbf{a}]_1^{\prime\prime}$ and $[j]_1$ to $P_3$.

    \item \textit{Array access.} $P_2$ and $P_3$ first reconstruct $h = I + r_{01}^{a}$ using their local values $[I]_2$ and $[I]_3$ and the received values $[j]_0$ and $[j]_1$. Based on $h$, they select values corresponding to $[\mathbf{a}]_0^{\prime\prime}$ and $[\mathbf{a}]_1^{\prime\prime}$. Next, $P_2$ sends $[s]_2 = [\mathbf{a}[h]]_0^{\prime\prime} + r_{23}^{a}$ to $P_1$ and $[s]_2^\prime = [\mathbf{a}[h]]_0^{\prime\prime} + r_{23}^{a} - r_{23}^{b}$ to $P_0$. Similarly, $P_3$ sends $[s]_3 = [\mathbf{a}[h]]_1^{\prime\prime} - r_{23}^{a}$ to $P_0$ and $[s]_3^\prime = [\mathbf{a}[h]]_1^{\prime\prime} - r_{23}^{a} + r_{23}^{b}$ to $P_1$.

    \item \textit{Share conversion} ($\langle \cdot \rangle$ to $\llbracket \cdot \rrbracket$). To generate the secret share $\llbracket \mathbf{a}[I] \rrbracket$, the parties set the following: $P_0$ assigns $[\mathbf{a}[I]]_0=[s]_3$ and $[\mathbf{a}[I]]_0^\prime=[s]_2^\prime$; $P_1$ assigns $[\mathbf{a}[I]]_1=[s]_2$ and $[\mathbf{a}[I]]_1^\prime=[s]_3^\prime$; $P_2$ assigns $[\mathbf{a}[I]]_2=[s]_2$ and $[\mathbf{a}[I]]_2^\prime=[s]_2^\prime$; $P_3$ assigns $[\mathbf{a}[I]]_3=[s]_3$ and $[\mathbf{a}[I]]_3^\prime=[s]_3^\prime$.
\end{enumerate}
The secure array access protocol $\Pi_{\mathsf{AA}}$ is elaborated in \textsc{Protocol} \ref{algo_aa}. 
Appendix~C provides the proof of Theorem \ref{theorem-aa}.

\begin{algorithm}
\SetKwInOut{Entities}{Entities}\SetKwInOut{Input}{Input}\SetKwInOut{Output}{Output}
\caption{Secure Array Access $\Pi_{\mathsf{AA}}$} 
\label{algo_aa} 

\Input{Secret shared array $\llbracket \mathbf{a} \rrbracket$ and secret shared index $\llbracket I \rrbracket$.} 
\Output{Secret shared $\llbracket \mathbf{a}[I] \rrbracket$.}
 
 Run the $\Pi_{\mathsf{SC}}$ to convert $\llbracket \mathbf{a} \rrbracket$ and $\llbracket I \rrbracket$ to $\langle \mathbf{a} \rangle$ and $\langle I \rangle$\;
 
 $P_0, P_1$ generate random numbers $r_{01}^{a}, \mathbf{r}_{01}^{b}$ and rotate $[\mathbf{a}]_i^\prime=[\mathbf{a}]_i \circlearrowright r_{01}^{a},i\in\{0,1\}$\;
 $P_0$ sets $[j]_0 = ([I]_0+r_{01}^{a}),[\mathbf{a}]_0^{\prime\prime}=[\mathbf{a}]_0^\prime \oplus \mathbf{r}_{01}^{b}$ and sends $[j]_0,[\mathbf{a}]_0^{\prime\prime}$ to $P_2$\;
 $P_1$ sets $[j]_1=([I]_1+r_{01}^{a}),[\mathbf{a}]_1^{\prime\prime} = [\mathbf{a}]_1^\prime \ominus \mathbf{r}_{01}^{b}$ and sends $[j]_1,[\mathbf{a}]_1^{\prime\prime}$ to $P_3$\;
 
 $P_2, P_3$ generate random numbers $r_{23}^{a}, r_{23}^{b}$ and reconstructs $h=I+r_{01}^{a}$ using the local $[I]_2,[I]_3$ and the received $[j]_0,[j]_1$\;
 $P_2$ sends $[s]_2=[\mathbf{a}[h]]_0^{\prime\prime}+r_{23}^{a}$ to $P_1$ and $[s]_2^\prime=[\mathbf{a}[h]]_0^{\prime\prime}+r_{23}^{a} - r_{23}^{b}$ to $P_0$\;
 $P_3$ sends $[s]_3=[\mathbf{a}[h]]_1^{\prime\prime}-r_{23}^{a}$ to $P_0$ and $[s]_3^\prime=[\mathbf{a}[h]]_1^{\prime\prime}-r_{23}^{a} + r_{23}^{b}$ to $P_1$\;
 
 $P_0$ sets $[\mathbf{a}[I]]_0=[s]_3, [\mathbf{a}[I]]_0^\prime=[s]_2^\prime$\;
 $P_1$ sets $[\mathbf{a}[I]]_1=[s]_2, [\mathbf{a}[I]]_1^\prime=[s]_3^\prime$\;
 $P_2$ sets $[\mathbf{a}[I]]_2=[s]_2, [\mathbf{a}[I]]_2^\prime=[s]_2^\prime$\;
 $P_3$ sets $[\mathbf{a}[I]]_3=[s]_3, [\mathbf{a}[I]]_3^\prime=[s]_3^\prime$\;
\end{algorithm}

\begin{theorem}
\label{theorem-aa}
\textit{Protocol $\Pi_{\mathsf{AA}}$ securely realizes $\mathcal{F}_{\mathsf{AA}}$ (see Fig.~11, Appendix~B) in the presence of one semi-honest corrupted party in the ($\mathcal{F}_{\mathsf{SC}}$)-hybrid model}.
\end{theorem}

\subsubsection{DReLU protocol} To implement the secure ReLU protocol, we first need to design the derivative protocol for the ReLU function, known as the DReLU protocol. If the input $x>0$, the protocol outputs 1; if the input $x \leq 0$, the protocol outputs 0. Our DReLU protocol, denoted as $\Pi_{\mathsf{DReLU}}$, is based on Bicoptor 2.0\cite{zhou2023bicoptor2} (Algorithm 7). We eliminated third parties from their protocol and made several modifications to $(2,4)$-RSS, as described in \textsc{Protocol} \ref{algo_drelu}. 
Simply porting the protocol to a four-party scenario raises privacy concerns. We introduce two random numbers $r_{01}^a$ and $r_{01}^b$ at $P_0$ and $P_1$ to mask $\text{DReLU}(x)$, as shown in lines 16-17 of \textsc{Protocol} \ref{algo_drelu}. Without these two random numbers, $P_2$ or $P_3$ could recover the exact value of $\text{DReLU}(x)$. Furthermore, unlike Bicoptor 2.0, our protocol does not require additional third-party assistance to recover $\mathbf{w}$ or generate masking random values during a preprocessing phase.
The non-interactive deterministic truncation protocol invoked by the DReLU protocol and the modulo-switch protocol $\Pi_{\mathsf{MS}}$ also adhere to Algorithm 5 and Algorithm 6 of Bicoptor 2.0. 
\begin{algorithm}
\SetKwInOut{Entities}{Entities}\SetKwInOut{Input}{Input}\SetKwInOut{Output}{Output}
\caption{DReLU $\Pi_{\mathsf{DReLU}}$} 
\label{algo_drelu} 

\Input{Secret shared $\llbracket x \rrbracket$.} 
\Output{Secret shared $\llbracket \text{DReLU}(x) \rrbracket$.}
 
 \tcp{Phase 1: Random masking and bit decomposition by $P_i$, where $i\in \{0,1\}$}
 $P_i$ use $seed_{01}$ to generate random bit $b$ and compute $[x]_i^\prime=(-1)^b \cdot [x]_i \bmod{2^\ell}$\;
 \For{$j\leftarrow 0$ \KwTo $\ell_x$} {
    $P_i$ compute $[\mathbf{u}_j]_i=[\overline{\sf{trc}}(x,j,\ell-\ell_x-j)]_i \bmod{2^{\ell_x}}$\;
 }
 \For{$k\leftarrow 0$ \KwTo $\ell_x-1$} {
    $P_i$ compute $[\mathbf{v}_k]_i=[\mathbf{u}_k]_i+[\mathbf{u}_{k+1}]_i -1\bmod{2^{\ell_x}}$\;
 }
 $P_i$ compute $[\mathbf{v}_{\ell_x}]_i=[\mathbf{u}_{\ell_x}]_i-1\bmod{2^{\ell_x}}$\;
 \For{$l\leftarrow 0$ \KwTo $\ell_x$} {$P_i$ run $\Pi_{\mathsf{MS}}$ switch $[\mathbf{v}_l]_i$ from the ring $\mathbb{Z}_{2^{\ell_x}}$ to $\mathbb{Z}_p$\;}
 $P_i$ shuffles $[\mathbf{v}]_i=\Pi_{shuffle}([\mathbf{v}]_i)$ and generates $\ell_x+1$ random numbers $\mathbf{r}_{01}=[r_{01}^0,\cdots,r_{01}^{\ell_x}]$ using $seed_{01}$, where $r_{01}^k \in \mathbb{Z}_p^*, \mathbb{Z}_p^* = \mathbb{Z}_p /\{0\},k\in[0,\ell_x]$\;
 \For{$j\leftarrow 0$ \KwTo $\ell_x$}{
    $P_i$ compute $[\mathbf{w}_j]_i=[v_j]_i\cdot \mathbf{r}_{01}[j] \bmod{p}$\;
 }
 $P_i$ sends $[\mathbf{w}]_i$ to $P_2$ (or $P_3$)\;

 \tcp{Phase 2: Zero testing by $P_2$ (or $P_3$)}
 $P_2$ (or $P_3$) picks a random value $r$ in $\mathbb{Z}_{2^\ell}$ and reconstructs $\mathbf{w}$ based on the received $[\mathbf{w}]_0,[\mathbf{w}]_1$. If there is 0 in $\mathbf{w}$, sets $\text{DReLU}(x)^{\prime\prime}=1$, otherwise, sets $\text{DReLU}(x)^{\prime\prime}=0$\;
 \tcp{Phase 3: DReLU output sharing}
 $P_2$ (or $P_3$) sends $[\text{DReLU}(x)^{\prime\prime}]_0=r\bmod{2^\ell}$ to $P_0$ and $[\text{DReLU}(x)^{\prime\prime}]_1=\text{DReLU}(x)^{\prime\prime}-r \bmod{2^\ell}$ to $P_1$\;
 $P_i$ compute $[\text{DReLU}(x)^\prime]_i=b+(1-2b)\cdot[\text{DReLU}(x)^{\prime\prime}]_i \bmod{2^\ell}$ and generate two random numbers $r_{01}^a, r_{01}^b$\;
 
 \tcp{Phase 4: Resharing into the (2,4)-RSS format}
 $P_0$ sets $[\text{DReLU}(x)]_0=[\text{DReLU}(x)^\prime]_0-r_{01}^a \bmod{2^\ell}$, $[\text{DReLU}(x)]_0^\prime=[\text{DReLU}(x)^\prime]_0-r_{01}^a+r_{01}^b \bmod{2^\ell}$\;
 $P_1$ sets $[\text{DReLU}(x)]_1=[\text{DReLU}(x)^\prime]_1+r_{01}^a \bmod{2^\ell}$, $[\text{DReLU}(x)]_1^\prime=[\text{DReLU}(x)^\prime]_1+r_{01}^a-r_{01}^b \bmod{2^\ell}$\;
 $P_0$ sends $[\text{DReLU}(x)]_0$ to $P_3$ and $[\text{DReLU}(x)]_0^\prime$ to $P_2$\; $P_1$ sends $[\text{DReLU}(x)]_1$ to $P_2$ and $[\text{DReLU}(x)]_1^\prime$ to $P_3$\;
 $P_2$ sets $[\text{DReLU}(x)]_2=[\text{DReLU}(x)]_1$ and $[\text{DReLU}(x)]_2^\prime=[\text{DReLU}(x)]_0^\prime$\;
 $P_3$ sets $[\text{DReLU}(x)]_3=[\text{DReLU}(x)]_0$ and $[\text{DReLU}(x)]_3^\prime=[\text{DReLU}(x)]_1^\prime$\;
\end{algorithm}

As shown in \textsc{Protocol}~\ref{algo_drelu}, our DReLU protocol consists of four major phases: (1) random masking and bit decomposition, (2) secure zero testing, (3) DReLU output sharing, and (4) resharing. The phases are described in detail as follows:
\begin{itemize}[leftmargin=*]
    \item \textbf{Phase 1: Random masking and bit decomposition.} Party $P_0$ and $P_1$ generates a random sign bit $b\in\{0,1\}$ using a shared seed and computes a masked version of the secret-shared input: $[x]_i^\prime=(-1)^b \cdot [x]_i \bmod{2^\ell}$, where $i\in \{0,1\}$. For $j=0$ to $\ell_x-1$, $P_0$ and $P_1$ locally compute the bit-decomposed shares $[\mathbf{u}_j]_i$ by performing truncation on the masked input. Then, for each bit index $k$, they compute: 
    \begin{align}
        [\mathbf{v}_k]_i = 
        \begin{cases}
        [\mathbf{u}_k]_i + [\mathbf{u}_{k+1}]_i - 1, & \text{for } k < \ell_x - 1, \\
        [\mathbf{u}_k]_i - 1, & \text{for } k = \ell_x - 1.
        \end{cases}
    \end{align}
    These values are converted from $\mathbb{Z}_{2^{\ell_x}}$ to $\mathbb{Z}_p$ using the protocol $\Pi_{\mathsf{MS}}$.
    \item \textbf{Phase 2: Secure zero testing.} Each $\mathbf{v}_k$ is masked with a random value $\mathbf{r}_{01}^{k} \in \mathbb{Z}_p^*$, and the masked shares $[\mathbf{w}]_i$ ($i\in\{0,1\}$) are sent to $P_2$ (or $P_3$), which reconstructs the value $\mathbf{w}=\{w_0,\cdots,w_k\cdots,w_{\ell_x-1}\}$. If any $w_k = 0$, then $\text{DReLU}(x)^{\prime\prime} = 1$; otherwise, $\text{DReLU}(x)^{\prime\prime} = 0$.
    \item \textbf{Phase 3: DReLU output sharing.} The value $\text{DReLU}(x)^{\prime\prime}$ is secret-shared using a new random mask value $r \in \mathbb{Z}_{2^\ell}$: $[\text{DReLU}(x)^{\prime\prime}]_0 = r$, $[\text{DReLU}(x)^{\prime\prime}]_1 = \text{DReLU}(x)^{\prime\prime} - r \bmod 2^\ell$. $P_i$ ($i\in\{0,1\}$) then computes: $[\text{DReLU}(x)]_i = b + (1 - 2b) \cdot [\text{DReLU}(x)^{\prime\prime}]_i \bmod 2^\ell$, to remove the effect of the sign bit $b$.
    \item \textbf{Phase 4: Resharing.} Finally, the shares are reshared to all four parties using the $(2,4)$-RSS scheme. This involves generating random values $r_{01}^a$ and $r_{01}^b$ from the seed and constructing replicated shares for $P_2$ and $P_3$.
\end{itemize}

Steps 1-13 of \textsc{Protocol} \ref{algo_drelu} mirror the DReLU protocol for Bicoptor 2.0. In Step 14, $(2,2)$-ASS of $\text{DReLU}(x)^{\prime\prime}$ is sent to $P_0$ and $P_1$. In Step 15, $P_0$ and $P_1$ locally remove the mask value $b$ from $\text{DReLU}(x)^{\prime\prime}$. It is easy to observe that $\text{DReLU}(x)=[\text{DReLU}(x)^{\prime}]_0+[\text{DReLU}(x)^{\prime}]_1$. Steps 16-21 convert $(2,2)$-ASS back to $(2,4)$-RSS. Steps 16-17 randomize $[\text{DReLU}(x)^{\prime}]_0$ and $[\text{DReLU}(x)^{\prime}]_1$ using random numbers $r_{01}^a$ and $r_{01}^b$, respectively. The use of $r_{01}^a$ aims to prevent $P_2$ (or $P_3$) from reconstructing $\text{DReLU}(x)$, given that $P_2$ (or $P_3$) has access to $[\text{DReLU}(x)^{\prime\prime}]_0$ and $[\text{DReLU}(x)^{\prime\prime}]_1$. 
Appendix~C contains the proof of Theorem \ref{theorem-drelu}.

\begin{theorem}
\label{theorem-drelu}
\textit{Protocol $\Pi_{\mathsf{DReLU}}$ securely realizes $\mathcal{F}_{\mathsf{DReLU}}$ (see Fig.~12, Appendix~B) in the presence of one semi-honest corrupted party in the ($\mathcal{F}_{\mathsf{MS}}$)-hybrid model}.
\end{theorem}

\subsubsection{ReLU protocol} 
$\text{ReLU}(x)$ can be computed as $\text{ReLU}(x)=x\cdot \text{DReLU}(x)$. Therefore, the result of $\text{ReLU}(x)$ can be computed by invoking protocols $\Pi_{\mathsf{Mult}}$ and $\Pi_{\mathsf{DReLU}}$. 
Appendix~C provides the proof of Theorem \ref{theorem-relu}.

\begin{theorem}
\label{theorem-relu} 
\textit{Protocol $\Pi_{\mathsf{ReLU}}$ securely realizes $\mathcal{F}_{\mathsf{ReLU}}$ (see Fig.~13, Appendix~B) in the presence of one semi-honest corrupted party in the ($\mathcal{F}_{\mathsf{Mult}}$, $\mathcal{F}_{\mathsf{DReLU}}$)-hybrid model}.
\end{theorem}

\subsubsection{Softmax protocol}
The softmax function normalizes a vector of values into a probability distribution, with each element ranging from 0 to 1 and summing to 1. For a $d$-dimensional input vector $\mathbf{x}=\{x_0,\cdots,x_d\}$, the output value of the $i$-th element is calculated as $\text{Softmax}(\mathbf{x}_i)=\frac{e^{\mathbf{x}_i}}{\sum_{j=1}^d \mathbf{x}_j}$, where $i\in \{1,2,\cdots,d\}$. To enable fast computation under ciphertext, most fixed-point schemes for computing the softmax activation function utilize approximation methods, such as SecureML\cite{mohassel2017secureml}. Our softmax protocol $\Pi_{\mathsf{Softmax}}$ is inspired by the approach proposed by Zheng et al.~\cite{zheng2023secure}, which approximates the softmax activation function using ordinary differential equations. The detailed process of approximation is described below:
\begin{itemize}[leftmargin=*]
    \item Initially, the multi-variable function $\text{Softmax}(\mathbf{x})$ is converted into a single-variable function $\mathbf{f}(\frac{i}{t})$, where $i$ is the independent variable and $t$ represents the number of loop iterations. For $i=0$, $\mathbf{f}(0)$ is defined as $[\frac{1}{d},\frac{1}{d},\cdots,\frac{1}{d}]$.
    \item Next, for $i\in [1,t]$, iteratively compute $\mathbf{f}(\frac{i}{t})$ as $\mathbf{f}(\frac{i-1}{t})+(\mathbf{x}-<\mathbf{x},\mathbf{f}(\frac{i-1}{t})>\mathbf{1 })\odot \mathbf{f}(\frac{i-1}{t})\cdot \frac{1}{t}$. Here, $<,>$ denotes the inner product of matrices (in the context of softmax, it represents a 1-dimensional matrix), and $\odot$ signifies element-wise matrix multiplication. Boldface letters denote matrices, and  $\mathbf{1}$ means that the matrix is filled with $1$.
    \item After $t$ iterations, the value of $\mathbf{f}(1)$ (i.e., $\mathbf{f}(\frac{t}{t})$), serves as the approximate output of $\text{Softmax}(\mathbf{x})$.
\end{itemize}

\begin{algorithm}
\SetKwInOut{Entities}{Entities}\SetKwInOut{Input}{Input}\SetKwInOut{Output}{Output}
\caption{Softmax $\Pi_{\mathsf{Softmax}}$} 
\label{algo_softmax} 
\SetAlgoLined
\Input{Secret shared $\llbracket \mathbf{x} \rrbracket$.} 
\Output{Secret shared $\llbracket \text{Softmax}(\mathbf{x}) \rrbracket$.}
 
 $P_i$ locally computes $[\mathbf{x}]_i=\frac{1}{t}\cdot [\mathbf{x}]_i$\;
 $P_0$ sets $([\mathbf{f}(0)]_0,[\mathbf{f})(0)]_0^\prime)=\{(\frac{1}{d},\frac{1}{d})\}^d$, $P_1$ sets $([\mathbf{f}(0)]_1,[\mathbf{f})(0)]_{1}^\prime)=\{(0,0)\}^d$, $P_2$ sets $([\mathbf{f}(0)]_2,[\mathbf{f})(0)]_2^\prime)=\{(0,\frac{1}{d})\}^d$, and $P_3$ sets $([\mathbf{f}(0)]_3,[\mathbf{f})(0)]_3^\prime)=\{(\frac{1}{d}, 0)\}^d$\;
 
 \For{$j\leftarrow 1$ \KwTo $t$} {
    $P_i$ computes $[a]_i=<\llbracket \mathbf{x} \rrbracket,\llbracket \mathbf{f}(\frac{j-1}{t})\rrbracket>$ and sets $[\mathbf{a}]_i=[a]_i \cdot \mathbf{1}$\;
    $P_i$ computes $[\mathbf{b}]_i=\Pi_{\mathsf{HadamardProd}}(\llbracket \mathbf{a} \rrbracket,\llbracket \mathbf{f}(\frac{j-1}{t})\rrbracket)$ and $[\mathbf{c}]_i=\Pi_{\mathsf{HadamardProd}}(\llbracket \mathbf{x} \rrbracket,\llbracket \mathbf{f}(\frac{j-1}{t})\rrbracket)$\;
    $P_i$ computes $[\mathbf{f}(\frac{j}{t})]_i=([\mathbf{c}]_i-[\mathbf{b}]_i)+[\mathbf{f}(\frac{j-1}{t})]_i$\;
 }
 $P_i$ sets $[\text{Softmax}(\mathbf{x})]_i=[\mathbf{f}(1)]_i$\;
\end{algorithm}

The detailed steps of $\Pi_{\mathsf{Softmax}}$ are outlined in \textsc{Protocol}~\ref{algo_softmax}. In Steps 1-2, each party $P_i, i\in \{0,1,2,3\}$ performs local computations of $\mathbf{x}\cdot \frac{1}{t}$ and initializes $\mathbf{f}(0)$. Steps 3-7 involve the computation of $\mathbf{f}(\frac{j}{t})$. In Step 4, the inner product of $\mathbf{x}$ and $\mathbf{f}(\frac{j-1}{t})$ is computed jointly for $j \in [1,t]$. Step 5 involves invoking $\Pi_{\mathsf{HadamardProd}}$ to compute the element-wise matrix multiplication. Subsequently, Step 6 computes the result of $\mathbf{f}(\frac{j}{t})$. Finally, Step 8 outputs the approximation of $\text{Softmax}(\mathbf{x})$. 
In this paper, as shown in TABLE~XIII (see Appendix~F), we choose $t = 8$.
The proof of Theorem \ref{theorem-softmax} is provided in Appendix~C.

\begin{theorem}
\label{theorem-softmax}
\textit{Protocol $\Pi_{\mathsf{Softmax}}$ securely realizes $\mathcal{F}_{\mathsf{Softmax}}$ (see Fig.~14, Appendix~B) in the presence of one semi-honest corrupted party in the ($\mathcal{F}_{\mathsf{Mult}}$, $\mathcal{F}_{\mathsf{HadamardProd}}$)-hybrid model}.
\end{theorem}

\subsubsection{Division protocol}
Performing efficient division operations under secret sharing is inherently challenging, and previous studies have employed the Newton-Raphson algorithm for approximate computation\cite{wagh2021falcon,tan2021cryptgpu,wang2023secgnn}. Given that $\frac{a}{b}=a\cdot \frac{1}{b}$, the emphasis lies in computing $\frac{1}{b}$. Computing $\frac{1}{b}$ requires iterative calculation of $y_{n+1}=y_n(2-by_n)$, eventually yielding an approximation $y_n\approx \frac{1}{b}$.
The key to the Newton-Raphson algorithm is the selection of initial values to minimize the number of iterations, and our division protocol $\Pi_{\mathsf{Div}}$ mirrors Falcon's\cite{wagh2021falcon}, adopting initial values consistent with those from previous works\cite{catrina2010secure,aliasgari2012secure}, i.e., $y_0=2.9142-2b$. 
In order to balance accuracy and computational cost, we adopt the same number of iterations as Falcon\cite{wagh2021falcon}, iterating up to $y_2$. Let $\beta_0=1-by_0, \beta_1=(1-by_0)^2=\beta_0^2$, then $\frac{a}{b}\approx a\cdot y_2=a\cdot y_0(1+\beta_0)(1+\beta_1)$.

In the Newton-Raphson algorithm, the divisor $b$ must satisfy $b\in[0.5,1)$. The range of divisors is computed using the bounding power protocol ($\Pi_{\mathsf{Pow}}$) from Falcon\cite{wagh2021falcon}. Division is performed according to Falcon's division protocol ($\Pi_{\mathsf{Div}}$). $\Pi_{\mathsf{Pow}}$ and $\Pi_{\mathsf{Div}}$ are adjusted to accommodate the $(2,4)$-RSS and replace the underlying $\Pi_{\mathsf{DReLU}}$ and $\Pi_{\mathsf{Mult}}$. The ideal functionality $\mathcal{F}_{\mathsf{Pow}}$ and $\mathcal{F}_{\mathsf{Div}}$ follows the prior work\cite{wagh2021falcon}.

\subsubsection{Inverse square root protocol} 
Following prior work~\cite{wagh2021falcon,wang2023secgnn}, we adopt the Newton-Raphson method to approximate $\frac{1}{\sqrt{x}}$ using the iterative formula: $y_{n+1} = \frac{y_n}{2}(3 - x y_n^2)$. 
The initial value $y_0$ is determined as $2^{-\lfloor \alpha / 2\rceil}$, where $\alpha$ is computed by $\Pi_{\mathsf{Pow}}$ and satisfies $2^\alpha \leq x < 2^{\alpha+1}$. Similar to Falcon\cite{wagh2021falcon}, 4 iterations are adequate to achieve the desired accuracy of the results (see Appendix~F for detailed analysis). The ideal functionality $\mathcal{F}_{\mathsf{InvSqrt}}$ follows the prior work\cite{wagh2021falcon}.

\section{Experimental Evaluation}
We design the following research questions to evaluate the privacy preservation, accuracy preservation, performance, and cost-effectiveness of \tool: 

\begin{itemize}[leftmargin=*]
    \item \textbf{Privacy preservation evaluation (RQ1):} Can \tool preserve privacy?
    \item \textbf{Accuracy preservation evaluation (RQ2):} Can \tool maintain accuracy while preserving privacy?
     \item \textbf{Performance evaluation (RQ3):} How is the efficiency and communication overhead of \tool compared with prior works? 
    \item \textbf{Cost-effectiveness evaluation (RQ4):} How is the estimated cost of \tool compared with prior works? 
    \item \textbf{Performance evaluation on individual modules (RQ5):} How do the individual modules perform regarding efficiency and communication overhead?
\end{itemize}

To address \textbf{RQ1}, we present the security proofs of the protocols proposed by \tool in Appendix~C. 
Following this, we evaluate the accuracy of \tool by comparing it with both plaintext models and SecGNN\footnote{We re-implemented SecGNN using our same backend since it was only used for local simulation and could not be run in a real environment.}\cite{wang2023secgnn}, thereby answering \textbf{RQ2}. To study \textbf{RQ3}, we compare the efficiency and communication overhead of \tool and SecGNN.
To explore \textbf{RQ4}, we estimate the cost of \tool using two pricing models (\textit{Plan A} and \textit{Plan B}, detailed in TABLE~\ref{table-cloud-pricing}) from three popular cloud platforms. To investigate \textbf{RQ5}, we measure the efficiency and communication overhead of \textit{secure array access protocol} and \textit{random neighbor padding method}, and evaluate the performance of each major sub-protocol.

\subsection{Experimental Setup}
All experiments were conducted on a computer with an Intel(R) Core(TM) i7-13700KF processor, 8 CPU cores, and 16 GB of RAM, running on Ubuntu 22.04 LTS. The experiments were performed over the ring $\mathbb{Z}_{2^{64}}$ with binary fractional precision $\ell_t=20$ and precision bits $\ell_x=31$. Additionally, the prime number $p=16381$ was selected for the relevant computations in the experiments. Our framework is implemented in C++ and the communication backend was constructed based on Falcon\cite{wagh2021falcon}. 
Our implementation is available at \url{https://github.com/CHEN-CONGCONG/Panther}.

\textbf{Communication Network}. We use 4 processes to simulate the 4 parties involved in the protocol computation. The \textsf{tc} (traffic control) tool was used to emulate different communication environments. Specifically, Local Area Network (LAN) was configured with 10 Gbps bandwidth and 0.3 ms round-trip latency; Wide Area Network (WAN) was configured with 400 Mbps bandwidth and 40 ms latency.

\textbf{Datasets}. We adopted three widely used graph datasets: Cora, CiteSeer, and PubMed, consistent with prior work~\cite{kipf2016semi,wang2023secgnn,Yuan2024PS-GNN}. Details are summarized in TABLE~\ref{table-datasets}. Following the setup in SecGNN~\cite{wang2023secgnn}, we partition each dataset into a training set with 40 labeled nodes per class, a validation set with 500 nodes, and a test set with 1,000 nodes.
\begin{table}[!ht]
\centering
\caption{Details of Cora, CiteSeer and PubMed Datasets.}
\label{table-datasets}
\begin{tabular}{cccccc}
\toprule
\textbf{Dataset} & \textbf{\#Nodes} & \textbf{\#Edges} & \textbf{\#Features} & \textbf{\#Classes} & \textbf{Max Degree} 
\\ 
\midrule
Cora      & 2,708  & 5,429  & 1,433     & 7       & 168    
\\ 
CiteSeer  & 3,327  & 4,732  & 3,703     & 6       & 99     
\\ 
PubMed    & 19,717 & 44,338 & 500      & 3       & 171     
\\ 
\bottomrule
\end{tabular}%
\end{table}

\textbf{Model hyperparameters}. We implemented a two-layer GCN following~\cite{kipf2016semi,wang2023secgnn}. The hidden layer dimension is set to 16. We used Stochastic Gradient Descent (SGD) with learning rate 0.15 and Kaiming initialization~\cite{he2015delving}. Inference follows the same hyperparameters. To emulate cloud inference outsourcing, we train GNN weights on plaintext for 2000 epochs and then convert them to ciphertext for secure inference.

\textbf{Baselines}. We compare our framework with two state-of-the-art methods: SecGNN~\cite{wang2023secgnn} and PS-GNN~\cite{Yuan2024PS-GNN}. As PS-GNN is tailored for inductive tasks, we restrict its evaluation to secure array access performance and inference efficiency. All other evaluations focus on the transductive setting of SecGNN. Each experiment was repeated five times, and we report averaged results.

\textbf{Other details}. Additional details on optimization techniques (e.g., use of the Eigen library, DReLU key bits optimization), as well as precomputation strategies for neighbor states initialization, are provided in Appendix~E.

\subsection{Accuracy Preservation Evaluation (RQ2)}
We evaluated the accuracy of \tool, as depicted in TABLE~\ref{table-inference-accuracy}. Experimental results demonstrate that \tool maintains a high accuracy comparable to plaintext, with no significant differences observed, and an error margin of less than 0.6\%. The accuracy of \tool is also similar to that of SecGNN~\cite{wang2023secgnn}, with a difference of less than 0.3\%. The disparity in output between the ciphertext and plaintext models in the table may arise from operations such as truncation.

\begin{table}[!ht]
\centering
\caption{Comparison of Accuracy of Plaintext, SecGNN, and \toolnott.}
\label{table-inference-accuracy}
\begin{threeparttable}
\begin{tabular}{cccc}
\toprule
\multirow{2}{*}{\textbf{Dataset}} & \multicolumn{1}{c}{\multirow{2}{*}{\textbf{Plaintext Accuracy (\%)}}} & \multicolumn{2}{c}{\textbf{Ciphertext Accuracy (\%)}} \\ \cline{3-4}
                                  & \multicolumn{1}{c}{}        & \textbf{\toolnott}    & \textbf{SecGNN}    \\
\midrule
Cora                              & 83.9                                                          & 83.3 ($-$0.06)                     & \textbf{83.5} ($-$0.04)     \\
CiteSeer                          & 71.8                                                            & \textbf{71.5} ($-$0.03)                     & 71.3 ($-$0.05)   \\
PubMed                            & 81.0                                                          & 80.9 ($-$0.01)                    & \textbf{81.0} ($-$0.00)  \\
\bottomrule
\end{tabular}%

\begin{tablenotes} 
    \footnotesize
    \item \textbf{Note:} Parenthesized values represent the accuracy loss relative to plaintext.
\end{tablenotes}
\end{threeparttable}
\end{table}

\subsection{Performance Evaluation (RQ3)}
We evaluated the online phase performance on the Cora, CiteSeer, and PubMed datasets in both LAN and WAN environments. Specifically, we measured the time and communication overhead for a single training round and for inference.
\subsubsection{Efficiency}
\begin{figure}[!h]
\centering
\subfloat[Comparison of time (LAN)]{\includegraphics[width=0.5\linewidth]{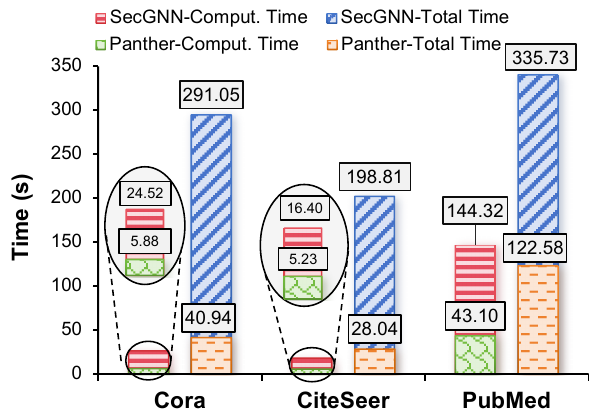}}
\hfil
\subfloat[Comparison of time (WAN)]{\includegraphics[width=0.5\linewidth]{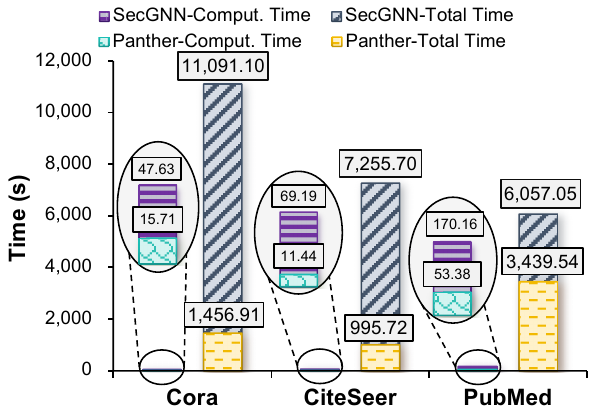}}
\caption{Efficiency comparison of SecGNN and \toolnott training phase.}
\label{fig-comparison-training}
\end{figure}

\textbf{Training efficiency.} 
Fig.~\ref{fig-comparison-training} presents a comparison of GNN training efficiency across diverse datasets in both LAN and WAN environments. The key metrics analyzed are total training time (\textit{Total Time}) and computation time (\textit{Comput. Time}). Total training time includes both computation time and communication time. As shown in Fig.~\ref{fig-comparison-training}, \tool demonstrates efficient GNN training in LAN environments with reduced computational effort. 
In WAN environments, although there is a decrease in overall efficiency, consistent trends are still observed. \tool consistently outperforms SecGNN in both total training time and computation time. Specifically, compared to SecGNN, \tool reduces total training time, including communication time, by an average of 75.28\%. Moreover, the computation time, excluding communication time, decreases by an average of 72.23\%. 

\begin{figure}[!ht]
\centering
\subfloat[Comparison of time (LAN)]{\includegraphics[width=0.5\linewidth]{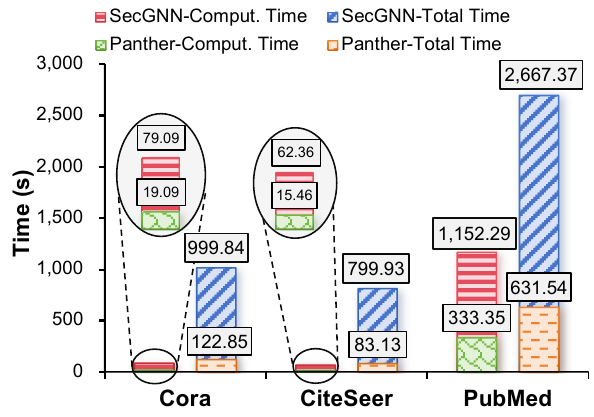}}
\hfil
\subfloat[Comparison of time (WAN)]{\includegraphics[width=0.5\linewidth]{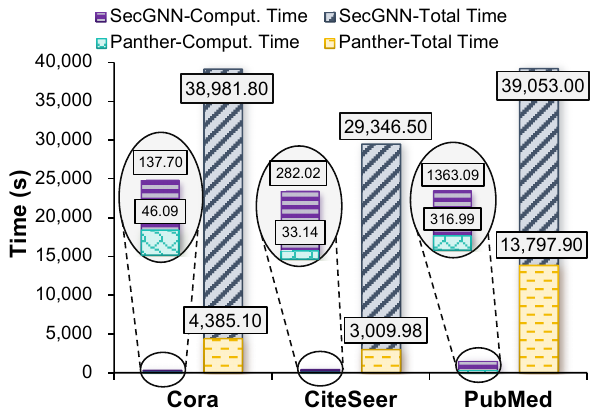}}
\caption{Efficiency comparison of SecGNN and \toolnott inference phase.}
\label{fig-comparison-inference}
\end{figure}

\textbf{Inference efficiency.}\label{sec-evaluation-inference-efficiency}
Experimental results shown in Fig.~\ref{fig-comparison-inference} show that \tool consistently surpasses SecGNN in both LAN and WAN environments. On average, \tool reduces the total inference time by 82.80\% compared to SecGNN, while also achieving a 50.26\% reduction in computation time. Even with increased communication latency in WAN settings, \tool retains its computational superiority. This improvement stems primarily from the \textit{random neighbor information padding} method and the \textit{secure array access protocol}.

\textbf{Summary.} Our experimental results show that \tool is more efficient than SecGNN in GNN training and inference across both LAN and WAN environments. Although SecGNN~\cite{wang2023secgnn} includes an offline phase to accelerate online computation, \tool still achieves higher overall efficiency.

\subsubsection{Communication Overhead}\label{sec-evaluation-overhead}
\begin{figure}[!ht]
\centering
\subfloat[Training phase]{\includegraphics[width=0.5\linewidth]{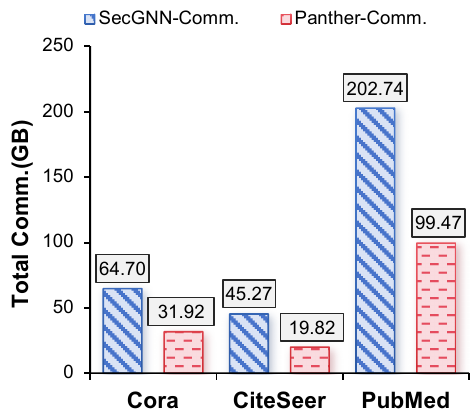}}
\hfil
\subfloat[Inference phase]{\includegraphics[width=0.5\linewidth]{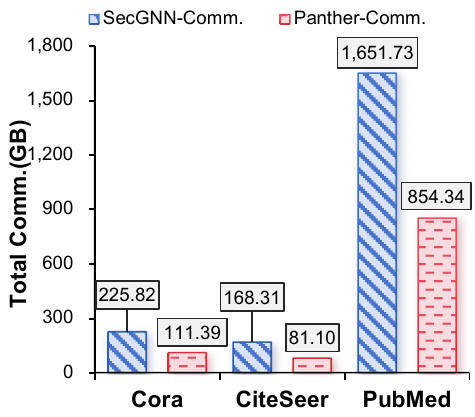}}
\caption{Comparison of communication overhead between SecGNN and \toolnott in the training and inference phases.}
\label{fig-comm-comparison}
\end{figure}

Fig.~\ref{fig-comm-comparison} show the communication overhead of \tool and SecGNN in the training and inference phases, respectively. The communication overhead is the same in both LAN and WAN environments. Compared to SecGNN\cite{wang2023secgnn}, the communication overhead in the training phase of \tool is reduced by 50.67\% to 56.22\%, and the communication overhead in the inference phase is reduced by approximately 48.28\% to 51.82\%.

\subsubsection{Inference Performance Comparison with PS-GNN}\label{sec-inference-psgnn-exp}
We further evaluate the inference performance of our framework by comparing it with PS-GNN~\cite{Yuan2024PS-GNN}. Since PS-GNN focuses on secure inference in inductive settings and does not support secure training, we only consider its inference performance. For a fair comparison, we re-implemented the inference functionality of PS-GNN within the \tool framework, adopting the same neighbor information protection strategy described in the original paper. For all experiments in this section, we randomly selected 20 target nodes from the Cora, CiteSeer, and PubMed datasets for inference evaluation. Consistent with PS-GNN, we set the number of sampled neighbors to $k_1=5$ and $k_2=10$ for the two aggregation layers. As shown in TABLE~\ref{table-psgnn-inference}, \tool consistently outperforms PS-GNN in inference efficiency under both LAN and WAN environments. Concretely, \tool achieves an average latency reduction of 89.62\% on LAN and 90.20\% on WAN, while maintaining almost identical communication overhead (with negligible average differences of 0.35\% and 0.34\%, respectively). This improvement is primarily attributed to the efficient secure array access protocol and optimized secure computation primitives in \tool, which effectively reduce computational and communication overhead during secure inference.
\begin{table}[t]
\centering
\caption{Comparison of Inference Performance Between \toolnott and PS-GNN~\cite{Yuan2024PS-GNN}.}
\label{table-psgnn-inference}
\renewcommand{\arraystretch}{1.1}
\resizebox{\linewidth}{!}{
\begin{tabular}{cccc|cc}
\toprule
                                 &                                    & \multicolumn{2}{c|}{\textbf{PS-GNN}}  & \multicolumn{2}{c}{\textbf{\toolnott}}   \\ \cline{3-6} 
\multirow{-2}{*}{\textbf{Network}} & \multirow{-2}{*}{\textbf{Dataset}} & \textbf{Time(s)} & \textbf{Comm.(MB)} & \textbf{Time(s)} & \textbf{Comm.(MB)} \\ \midrule
                                                          & Cora                               & 13.7191           & 124.2129            & 0.9703          & 123.2781         \\
                                                          & CiteSeer                           & 11.8802          & 314.0314         & 0.9686          & 313.3214        \\
   \multirow{-3}{*}{LAN}            & PubMed                             & 5.6089          & 44.8969            & 0.8926          & 44.8700            \\ \hline
                                                          & Cora                               & 466.8695         & 124.2129            & 31.0211         & 123.2781           \\
                                                          & CiteSeer                           & 400.2271         & 314.0314           & 29.8647         & 313.3214            \\
  \multirow{-3}{*}{WAN}                                 & PubMed                             & 191.5377         & 44.8969            & 29.3092         & 44.8700            \\
 
\bottomrule
\end{tabular}
}
\end{table}

\subsection{Cost-Effectiveness Evaluation (RQ4)}\label{sec-cost-effectiveness-evaluation}
Based on the financial cost data in TABLE~\ref{table-cloud-pricing}, TABLE~\ref{table-cost} presents the estimated training and inference financial costs of \tool and SecGNN. We assume that each instance has a 100GB disk and operates in the WAN environment. 
As shown in TABLE~\ref{table-cost}, \tool significantly reduces financial costs across all platforms compared to SecGNN. On average, \tool reduces training costs by 55.05\% under \textit{Plan A} and 59.00\% under \textit{Plan B}. Similarly, \tool reduces inference costs by 53.38\% and 56.75\% under the same pricing models.

\begin{table*}[!hbpt]
\centering
\renewcommand{\arraystretch}{1.1}
\caption{Estimated Financial Cost Comparison between SecGNN and \toolnott in the Training and Inference Phases.}
\label{table-cost}
\resizebox{\linewidth}{!}{%
\begin{threeparttable}
\begin{tabular}{ccccccccccc}
\toprule
\multirow{2}{*}{\textbf{Phase}}       & \multirow{2}{*}{\textbf{Dataset}} & \multirow{2}{*}{\textbf{Framework (\#Instances)}}  & \multirow{2}{*}{\textbf{\begin{tabular}[c]{@{}c@{}}Data \\ Transfer (GB)\end{tabular}}} & \multirow{2}{*}{\textbf{\begin{tabular}[c]{@{}c@{}}Total \\ Time (hour)\end{tabular}}} & \multicolumn{3}{c}{\textbf{Financial Cost (Plan A)$^1$}}  & \multicolumn{3}{c}{\textbf{Financial Cost (Plan B)$^2$}}        \\ \cline{6-11}
&    &  &  &   & \textbf{GCP} & \textbf{Azure} & \textbf{AWS} & \textbf{GCP} & \textbf{Azure} & \textbf{AWS} \\
\midrule
\multirow{7}{*}{\begin{tabular}[c]{@{}c@{}}Training\\ phase\\ (1 epoch)\end{tabular}} & \multirow{2}{*}{Cora}  & \toolnott (4)   & 15.96   & 0.40  & \textbf{\$1.67}  & \textbf{\$2.90}  & \textbf{\$1.85} & \textbf{\$1.52}  & \textbf{\$2.81}  & \textbf{\$1.74}  \\
                &     & SecGNN (3)   & 32.35   & 3.08  & \$4.76   & \$5.68  & \$5.24  & \$3.96   & \$5.14  & \$4.58  \\ 
 \cline{2-11}
 & \multirow{2}{*}{CiteSeer}  & \toolnott (4)  & 9.91  & 0.28   & \textbf{\$1.06}  & \textbf{\$2.32}  & \textbf{\$1.18}  & \textbf{\$0.96}  & \textbf{\$2.26}  & \textbf{\$1.10}   \\
    &      & SecGNN (3)   & 22.63    & 2.02    & \$3.24  & \$4.29  & \$3.56 & \$2.71  & \$3.94  & \$3.13 \\ \cline{2-11} 
& \multirow{2}{*}{PubMed}   & \toolnott (4)   & 49.74   & 0.96   & \textbf{\$4.88}  & \textbf{\$6.03}  & \textbf{\$5.44} & \textbf{\$4.55}  & \textbf{\$5.81}  & \textbf{\$5.17}    \\
&      & SecGNN (3)     & 101.37  & 1.68   & \$9.30   & \$10.39 & \$10.40 & \$8.86   & \$10.10 & \$10.04   \\ 
\hline
& \multicolumn{2}{c}{\textbf{Average Reduction ($\downarrow$):}}   & 
\multicolumn{1}{c}{\textbf{52.61\%}} & 
\multicolumn{1}{c}{\textbf{72.12\%}} & \multicolumn{1}{c}{\textbf{59.87\%}} & \multicolumn{1}{c}{\textbf{45.59\%}} & \multicolumn{1}{c}{\textbf{59.69\%}} & \multicolumn{1}{c}{\textbf{58.20\%}} & \multicolumn{1}{c}{\textbf{43.53\%}} & \multicolumn{1}{c}{\textbf{58.42\%}} \\ \hline
\multirow{7}{*}{\begin{tabular}[c]{@{}c@{}}Inference\\ phase\end{tabular}}  & \multirow{2}{*}{Cora}  & \toolnott (4)   & 55.69   & 1.22  & \textbf{\$5.61}  & \textbf{\$6.71}  & \textbf{\$6.24} & \textbf{\$5.18}  & \textbf{\$6.42}  & \textbf{\$5.90}    \\
     &    & SecGNN (3)  & 112.91  & 10.83   & \$16.64 & \$16.59   & \$18.30 & \$13.83 & \$14.70   & \$15.98 \\ 
\cline{2-11}   & \multirow{2}{*}{CiteSeer}   & \toolnott (4)  & 40.55  & 0.84   & \textbf{\$4.04}  & \textbf{\$5.20}   & \textbf{\$4.50} & \textbf{\$3.74}  & \textbf{\$5.01}   & \textbf{\$4.26}   \\
  &    & SecGNN (3)  & 84.16 & 8.15  & \$12.46  & \$12.74    & \$13.70 & \$10.34  & \$11.32    & \$11.96 \\ 
\cline{2-11} & \multirow{2}{*}{PubMed}   & \toolnott (4) & 427.17  & 3.83  & \textbf{\$37.77}  & \textbf{\$38.43}   & \textbf{\$42.29} & \textbf{\$36.44}  & \textbf{\$37.54}   & \textbf{\$41.20}     \\
                         &          & SecGNN (3) & 825.86  & 10.85 & \$73.69 & \$73.63    & \$82.48   & \$70.87 & \$71.74    & \$80.16     \\
\hline
& \multicolumn{2}{c}{\textbf{Average Reduction ($\downarrow$):}}   & 
\multicolumn{1}{c}{\textbf{50.26\%}} & 
\multicolumn{1}{c}{\textbf{81.05\%}} & \multicolumn{1}{c}{\textbf{60.88\%}} & \multicolumn{1}{c}{\textbf{55.52\%}} & \multicolumn{1}{c}{\textbf{60.59\%}} & \multicolumn{1}{c}{\textbf{58.30\%}} & \multicolumn{1}{c}{\textbf{53.25\%}} & \multicolumn{1}{c}{\textbf{58.69\%}}\\ 
\bottomrule
\end{tabular}%

\begin{tablenotes} 
    \footnotesize
    \item \textbf{Note:} Cost $=$ \#Instances $\times$ \texttt{Instance Cost} $\times$ total time $+$ \texttt{Data Transfer Cost} $\times$ outbound data transfer $+$ disk size $\times$ \texttt{Disk Storage Cost}.
    \item $^1$: \textit{Plan A} refers to the on-demand pricing model. $^2$: \textit{Plan B} refers to the 1-year resource commitment pricing model.

\end{tablenotes}
\end{threeparttable}
}
\end{table*}

\subsection{Performance Evaluation on Individual Modules (RQ5)}
\subsubsection{Comparison of neighbor information padding method}
\begin{figure}[!ht]
\centering
\subfloat[Time (Cora)]{\includegraphics[width=0.5\linewidth]{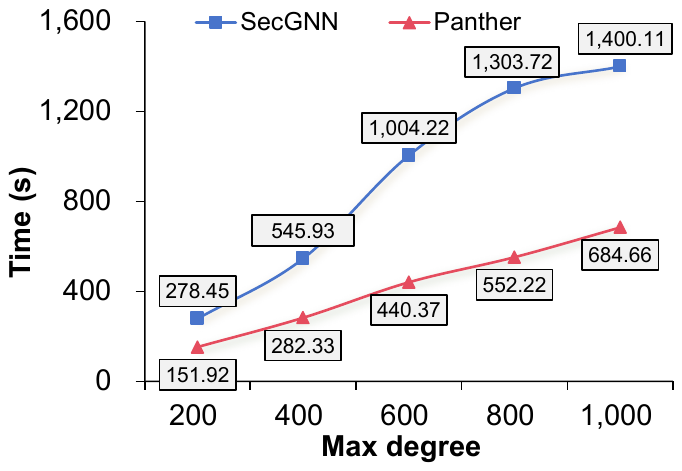}}
\subfloat[Comm. overhead (Cora)]{\includegraphics[width=0.5\linewidth]{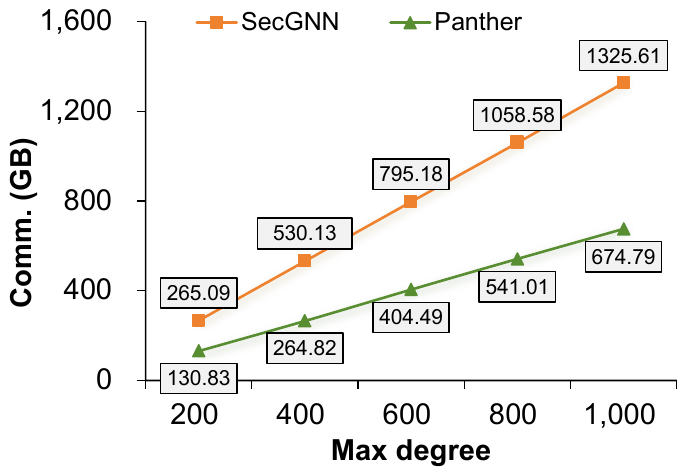}}
\hfil
\subfloat[Time (CiteSeer)]
{\includegraphics[width=0.5\linewidth]{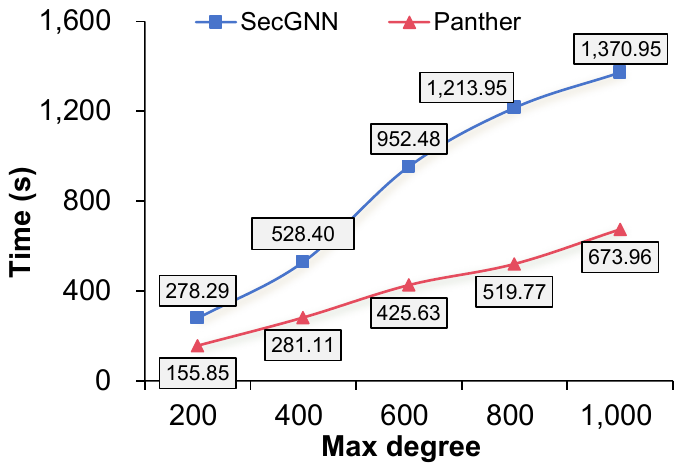}}
\subfloat[Comm. overhead (CiteSeer)]
{\includegraphics[width=0.5\linewidth]{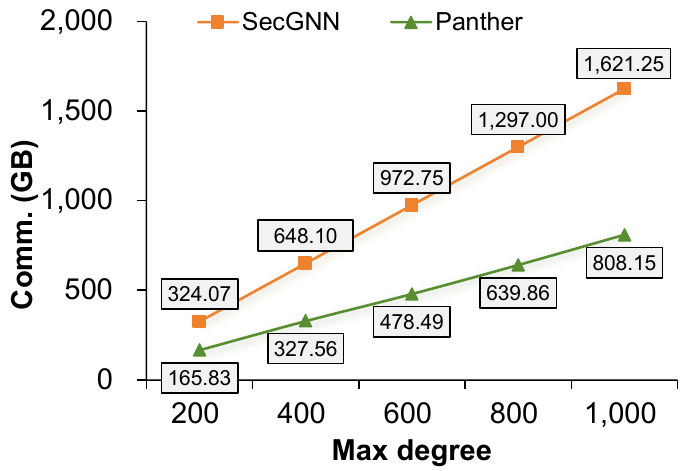}}
\hfil
\subfloat[Time (PubMed)]
{\includegraphics[width=0.5\linewidth]{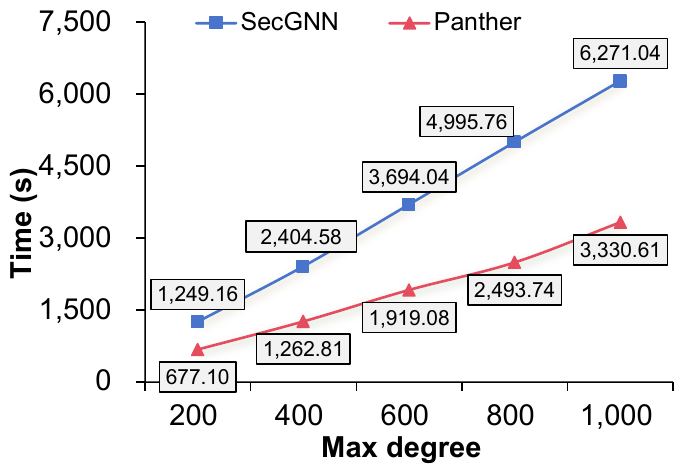}}
\subfloat[Comm. overhead (PubMed)]
{\includegraphics[width=0.5\linewidth]{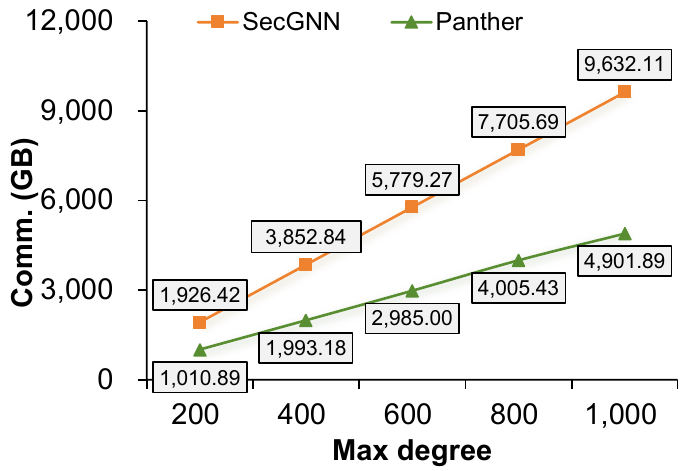}}
\caption{Performance comparison for neighbor information padding methods. Subfigures (a), (c), and (e) compare efficiency for Cora, CiteSeer, and PubMed datasets, respectively, while subfigures (b), (d), and (f) compare communication overhead for the same datasets.}
\label{fig-neigh-struc}
\end{figure}

To evaluate our \textit{random neighbor information padding} method, we compared \tool with SecGNN during the inference phase on 1,000 test samples from the Cora, CiteSeer, and PubMed datasets in a LAN environment. Additionally, we vary \emph{max degree} to 200, 400, 600, 800, and 1,000 to simulate scenarios where the degrees of nodes differ significantly. The detailed performance comparison results are presented in Fig.~\ref{fig-neigh-struc}. We observe that \tool exhibits approximately a \textit{twofold} improvement in both efficiency and communication overhead, with average reductions of 49.27\% and 50.07\%, respectively, offering significant advantages in scenarios characterized by larger maximum degrees.

\subsubsection{Comparison of secure array access protocol}\label{sec-aa-performence-evaluation}
We conducted one round of \textit{secure array access protocol} in the LAN environment on the Cora, CiteSeer, and PubMed datasets. TABLE~\ref{table-aa-comparison} compares the efficiency and communication overhead of \tool with those of the protocols proposed by SecGNN\cite{wang2023secgnn} and PS-GNN\cite{Yuan2024PS-GNN}. It is noticeable that the communication overhead of \tool slightly surpasses that of SecGNN and PS-GNN. This disparity arises because \tool engages four parties, whereas SecGNN and PS-GNN engages only three. Despite involving more parties, \tool reduces the overall time by an average of 36.74\% compared to SecGNN and PS-GNN. Additionally, its computation time decreases by an average of 45.80\% when compared to SecGNN and PS-GNN. This observation is consistent with the computational models: \tool uses an asynchronous mode, whereas SecGNN and PS-GNN use synchronous modes. \tool's asynchronous approach is particularly advantageous for privacy-preserving GNNs, as it excels when the secure array access protocol is invoked multiple times during training and inference.

\begin{table}[!ht]
\centering
\caption{Performance Comparison of Secure Array Access Protocol.}
\renewcommand{\arraystretch}{1.2}
\label{table-aa-comparison}
\resizebox{\linewidth}{!}{%
\begin{tabular}{ccccccc}
\toprule
\multirow{2}{*}{\textbf{Dataset}} & 
\multirow{2}{*}{\textbf{Framework}} & 
\textbf{Total} & 
\textbf{Comp.} & 
\textbf{Comm.}  &
\multicolumn{2}{c}{\textbf{Time
 Reduction ($\downarrow$)}} \\
\cline{6-7}
   &    & \textbf{Time (ms)}  &  \textbf{Time (ms)} &  \textbf{(MB)} & \textbf{Total}  & \textbf{Comp.}  \\
\midrule
\multirow{2}{*}{Cora}     & \toolnott            & 87.64                            & 40.88                            & 121.36   &  \multirow{2}{*}{\textbf{37.04\%}} & \multirow{2}{*}{\textbf{50.68\%}}     \\
                          & SecGNN, PS-GNN            & 139.20                           & 82.89                            & 121.31            \\ \hline
\multirow{2}{*}{CiteSeer} & \toolnott            & 298.47                           & 151.07                            & 383.49   &   \multirow{2}{*}{\textbf{34.18\%}} & \multirow{2}{*}{\textbf{44.76\%}}     \\
                          & SecGNN, PS-GNN             & 453.48                           & 273.45                           & 383.38             \\ \hline
\multirow{2}{*}{PubMed}   & \toolnott            & 230.00                           & 118.69                           & 308.11   &   \multirow{2}{*}{\textbf{36.74\%}} & \multirow{2}{*}{\textbf{45.80\%}}     \\
                          & SecGNN, PS-GNN             & 363.61                           & 218.98                           & 308.09            \\
\bottomrule
\end{tabular}%
}
\end{table}

\subsubsection{Comparison of sub-protocols}
To analyze the communication complexity introduced by \tool's four-party protocol, we present a detailed breakdown of time and communication overhead per protocol in TABLE~\ref{table-subprotocol-comparison}. The breakdown includes the core operations of GNN training and inference: $\Pi_\mathsf{AA}$, $\Pi_\mathsf{ReLU}$, $\Pi_\mathsf{Division}$, $\Pi_\mathsf{MatMul}$, $\Pi_\mathsf{HadamardProd}$, $\Pi_\mathsf{Softmax}$, and $\Pi_\mathsf{InvSqrt}$, across three datasets and different network settings. Despite the increased number of parties involved, \tool maintains low or comparable communication overhead through its carefully crafted protocols, thereby reducing the overall system overhead in both computational and communication. This design also leads to better scalability, especially in high-latency WAN environments.

\begin{table*}[!ht]
\centering
\caption{Time and Communication Breakdown of Each Protocol in \toolnott and Baseline Frameworks across Three Datasets and Different Network Environments (LAN/WAN).}
\renewcommand{\arraystretch}{1.1}
\label{table-subprotocol-comparison}
\resizebox{\linewidth}{!}{%
\begin{threeparttable}
\begin{tabular}{ccc|cc|cc|cc|cc|cc|cc|cc}
\toprule
\multirow{2}{*}{\textbf{Dataset}} & \multirow{2}{*}{\textbf{Network}} & \multirow{2}{*}{\textbf{Framework}} & \multicolumn{2}{c|}{$\Pi_\mathsf{AA}$$^1$}          & \multicolumn{2}{c|}{$\Pi_\mathsf{ReLU}$}        & \multicolumn{2}{c|}{$\Pi_\mathsf{Division}$}    & \multicolumn{2}{c|}{$\Pi_\mathsf{MatMul}$$^2$}      & \multicolumn{2}{c|}{$\Pi_\mathsf{HadamardProd}$}     & \multicolumn{2}{c|}{$\Pi_\mathsf{Softmax}$$^3$}     & \multicolumn{2}{c}{$\Pi_\mathsf{InvSqrt}$}     \\ \cline{4-17} 
                                  &                                   &                                     & \textbf{Time} & \textbf{Comm.} & \textbf{Time} & \textbf{Comm.} & \textbf{Time} & \textbf{Comm.} & \textbf{Time} & \textbf{Comm.} & \textbf{Time} & \textbf{Comm.} & \textbf{Time} & \textbf{Comm.} & \textbf{Time} & \textbf{Comm.} \\ \midrule
\multirow{4}{*}{Cora}             & \multirow{2}{*}{LAN}              & Panther                             & 87.64            & 121.36              & 4.03               & 0.47                & 22.92              & 3.18                & 0.27               & 0.002                & 0.03               & 0.18                & 21.32              & 0.02                & 26.44              & 4.43              \\
                                  &                                   & SecGNN                              & 139.20           & 121.31              & 27.84              & 63.28               & 37.21              & 3.13                & 1.95               & 2.91                & 0.89               & 0.09                & 586.45             & 0.09                & 104.76             & 8.33              \\ \cline{2-17} 
                                  & \multirow{2}{*}{WAN}              & Panther                             & 1,274.19          & 121.36              & 83.37              & 0.47                & 505.94             & 3.26                & 0.17               & 0.002                & 0.25               & 0.18                & 741.52             & 0.02                & 625.80             & 4.43              \\
                                  &                                   & SecGNN                              & 1,423.09          & 121.31              & 682.31             & 63.28               & 1,451.14            & 3.13                & 42.61              & 2.91                & 41.06              & 0.09                & 21,658.10           & 0.09                & 3,841.91            & 8.33              \\ \hline
\multirow{4}{*}{CiteSeer}         & \multirow{2}{*}{LAN}              & Panther                             & 298.47           & 383.49              & 5.91               & 1.22                & 36.40              & 8.22                & 0.36               & 0.002                & 0.07               & 0.46                & 17.60              & 0.02                & 40.80              & 11.46             \\
                                  &                                   & SecGNN                              & 453.48           & 383.38              & 368.46             & 163.51              & 38.60              & 8.10                & 6.03               & 7.52                & 0.93               & 0.23                & 497.25             & 0.07                & 103.85             & 21.52             \\ \cline{2-17} 
                                  & \multirow{2}{*}{WAN}              & Panther                             & 4,047.33          & 383.49              & 85.70              & 1.22                & 518.27             & 8.22                & 0.50               & 0.002                & 0.08               & 0.46                & 665.17             & 0.02                & 645.42             & 11.46             \\
                                  &                                   & SecGNN                              & 4,324.19          & 383.38              & 1758.27            & 163.51              & 1,456.63            & 8.10                & 80.67              & 7.52                & 40.78              & 0.23                & 18,254.52           & 0.07                & 3,890.45            & 21.52             \\ \hline
\multirow{4}{*}{PubMed}           & \multirow{2}{*}{LAN}              & Panther                             & 230.00           & 308.11              & 2.99               & 0.16                & 15.76              & 1.11                & 0.05               & 0.002                & 0.02               & 0.06                & 17.19              & 0.01                & 20.15              & 1.55              \\
                                  &                                   & SecGNN                              & 363.61           & 308.09              & 9.63               & 22.08               & 39.79              & 1.09                & 1.40               & 1.02                & 0.87               & 0.03                & 213.77             & 0.03                & 106.20             & 2.91              \\ \cline{2-17} 
                                  & \multirow{2}{*}{WAN}              & Panther                             & 3,252.85          & 308.11              & 82.24              & 0.16                & 500.64             & 1.14                & 0.06               & 0.002                & 0.02               & 0.06                & 663.32             & 0.01                & 624.93             & 1.55              \\
                                  &                                   & SecGNN                              & 3,496.54          & 308.09              & 266.16             & 22.08               & 1,449.02            & 1.09                & 41.39              & 1.02                & 41.01              & 0.03                & 7,894.22            & 0.03                & 3,872.04            & 2.91              \\ 
\bottomrule
\end{tabular}
\begin{tablenotes} 
    \footnotesize
    \item \textbf{Note:} All protocols are evaluated for a single invocation. \textit{Time} is measured in milliseconds (ms), and \textit{Comm.} is measured in megabytes (MB). Unless otherwise specified, the input size is set to $1 \times D_{in}$.
    \item $^1$: The input array size for $\Pi_\mathsf{AA}$ is $N \times D_{in}$. 
      $^2$: For $\Pi_\mathsf{MatMul}$, the matrix sizes are $1  \times D_{in}$ and $D_{in}\times 16$. 
     $^3$: The input size for $\Pi_\mathsf{Softmax}$ is the number of classes, $D_{out}$.
\end{tablenotes}
\end{threeparttable}
}
\end{table*}

\subsection{Scalability Evaluation}
\label{sec:appendix-scalability}
To evaluate the scalability of \tool, we conducted a focused experiment targeting the core cryptographic bottleneck: the \textit{secure array access protocol}. This protocol constitutes a major portion of computational and communication overhead in privacy-preserving GNN workflows, particularly during neighbor states aggregation.

We benchmarked the secure array access protocol with 100 runs under both LAN and WAN settings across five datasets of increasing scale. For each dataset, the array length was set to the number of nodes multiplied by the feature dimension. For smaller datasets (i.e., Cora, CiteSeer, and PubMed), we used all nodes, and their feature sizes are consistent with those listed in TABLE~\ref{table-datasets}. For large graphs, we utilized the entire ogbn-arxiv dataset, which contains 169,343 nodes with a feature dimension of 128. Due to hardware memory limitations, we randomly sampled one million nodes from the ogbn-products dataset, using a feature dimension of 100.

\textbf{Observation.} As shown in TABLE~\ref{table-scalability}, \tool significantly reduces execution time across all datasets and network settings, with minimal or negligible increase in communication overhead. Specifically, in the LAN setting, \tool achieves a 36.2\% to 40.25\% reduction in time on small to medium datasets (e.g., Cora, CiteSeer, and PubMed), and up to 38.81\% reduction on ogbn-arxiv. Even for large graphs such as ogbn-products, \tool still offers a notable 13.5\% improvement in time. In the WAN setting, where latency becomes a dominant factor, \tool shows 5.26\% to 18.45\% time reduction, with greater improvements observed on larger datasets, such as ogbn-products. Meanwhile, the communication overhead remains nearly unchanged, with differences consistently below 0.04\%, indicating that the performance gains of \tool do not come at the cost of increased bandwidth consumption.
\begin{table}[!ht]
\caption{Scalability Comparison of Secure Array Access Protocol}
\label{table-scalability}
\renewcommand{\arraystretch}{1.1}
\resizebox{\linewidth}{!}{
\begin{threeparttable}
\begin{tabular}{cccc|cc}
\toprule
                                 &                                    & \multicolumn{2}{c|}{\textbf{\toolnott}}  & \multicolumn{2}{c}{\textbf{SecGNN}}   \\ \cline{3-6} 
\multirow{-2}{*}{\textbf{Network}} & \multirow{-2}{*}{\textbf{Dataset}} & \textbf{Time(s)} & \textbf{Comm.(GB)} & \textbf{Time(s)} & \textbf{Comm.(GB)} \\ \midrule
                                                          & Cora                               & 8.6670           & 11.8513            & 14.5043          & 11.8469            \\
                                                          & CiteSeer                           & 29.1731          & 37.4504            & 45.7697          & 37.4391            \\
                                                          & PubMed                             & 23.4187          & 30.0888            & 36.7066          & 0.0873            \\
                                                          & ogbn-arxiv                         & 50.8360          & 66.1504            & 83.0817          & 66.1500            \\
\multirow{-5}{*}{LAN}                                     & Ogbn-Products$^*$                  & 1,091.4120       & 305.1767           & 1,261.7920       & 305.1758           \\ \hline
                                                          & Cora                               & 130.6898         & 11.8513            & 144.7944         & 11.8469            \\
                                                          & CiteSeer                           & 408.8210         & 37.4504            & 434.8736         & 37.4391            \\
                                                          & PubMed                             & 328.8534         & 30.0888            & 352.6492         & 30.0873            \\
                                                          & ogbn-arxiv                         & 719.8794         & 66.1504            & 759.8320         & 66.1500            \\
\multirow{-5}{*}{WAN}                                     & ogbn-products$^*$                   & 3,525.4180       & 305.1767           & 4,322.8120       & 305.1758         \\
\bottomrule
\end{tabular}
\begin{tablenotes} 
    \footnotesize
    \item \textbf{$^*$:} Ogbn-products randomly sampled one million nodes, due to hardware memory constraints.
\end{tablenotes}
\end{threeparttable}
}
\end{table}

Taken together, these findings indicate that \tool offers more substantial benefits on large graphs, especially when deployed in high-latency environments. Since privacy-preserving GNNs typically require encrypted processing of graph structural information such as neighbor IDs and edge weights, the overhead incurred by secure protocols becomes even more pronounced as graph size increases. Therefore, \tool's efficient design makes it particularly suitable for scalable, real-world privacy-preserving GNN training and inference in cloud environments.
\section{Related Work}

Recent research in privacy-preserving GNNs has focused on three main approaches: FL, DP, and cryptography. Our work falls into the cryptography domain, aiming to overcome the performance and cost limitations of prior methods.

\subsection{Privacy-Preserving GNNs via FL and DP}
FL is a prominent framework for training GNNs on decentralized data. Within this paradigm, various approaches to model aggregation have been explored. For instance, Chen et al.~\cite{chen2022vertically} proposed a method for vertically partitioned data that combines FL with secret sharing and differential privacy. In contrast, Pei et al.~\cite{pei2021decentralized} eliminated the central server entirely, using the Diffie-Hellman key exchange for secure, peer-to-peer parameter aggregation. However, such methods still pose privacy risks~\cite{wang2019beyond,zhu2019deep}, as sensitive information can be inferred from shared model updates.

Other works leverage DP to inject statistical noise, protecting data at the local level before it is shared. Daigavane et al.~\cite{daigavane2022node} and Mueller et al.~\cite{mueller2022differentially} focus on node-level and graph-level DP-based training, respectively. LGA-PGNN~\cite{pei2023privacy-lga-pgnn} enforces Local DP (LDP) on decentralized local graphs and uses a local graph augmentation technique to enhance the model's expressiveness, particularly for low-degree vertices. Similarly, \textsc{Blink}~\cite{zhu2023blink} introduces a novel link-level LDP mechanism that separates privacy budgets for links and degrees, using Bayesian estimation on the server side to denoise the graph structure and improve model accuracy. While these DP-based frameworks provide strong privacy guarantees for GNNs, they often face trade-offs between privacy and utility, where the injected noise can compromise final model accuracy.

\subsection{Privacy-Preserving GNNs via Cryptography}
Cryptography-based approaches offer a different paradigm, enabling computation on encrypted data to provide strong security without sacrificing model accuracy. \textbf{HE} allows for direct computation on ciphertexts. CryptoGCN~\cite{ran2022cryptogcn} and LinGCN~\cite{peng2024lingcn} are HE-based frameworks for secure GCN inference. They leverage techniques like matrix sparsity, pruning, and polynomial approximations to reduce the computational depth and overhead of HE operations. Penguin~\cite{ran2024penguin} further enhances efficiency using packed HE. Despite its effectiveness for inference, HE suffers from two key limitations: it typically leaves graph structures and model parameters unprotected, and its substantial computational overhead makes it infeasible for complex GNN training.

\textbf{SS} is a more lightweight cryptographic alternative. OblivGNN~\cite{xu2024oblivgnn} utilizes Function Secret Sharing (FSS) to build a highly efficient and oblivious inference service that supports both static (transductive) and dynamic (inductive) graphs, hiding data access patterns from the servers. However, its scope is limited to inference and does not address secure training. Furthermore, it involves a substantial offline overhead. In contrast, SecGNN~\cite{wang2023secgnn} is SS-based framework that support both secure training and inference. While it represents the state-of-the-art for end-to-end privacy-preserving GNN services, its practicality is limited by significant drawbacks. The method's reliance on padding neighbor information to the maximum degree is highly inefficient for large sparse graphs. Moreover, its synchronous secure array access protocol incurs substantial latency, resulting in high operational costs in cloud environments. Similarly, PS-GNN~\cite{Yuan2024PS-GNN}, another SS-based secure inference framework, which is tailored for GraphSAGE and supports inductive tasks, suffers from the same vulnerability in its secure array access protocol.

Our work, \tool, addresses these limitations by introducing an asynchronous four-party computation protocol and a randomized neighbor information padding method. These enhancements significantly reduce computational and communication overhead, making privacy-preserving GNN training and inference more practical and cost-effective. As a result, \tool lays a strong foundation for the broader adoption of secure GNN applications in real-world settings.

\section{Discussions}
\subsection{Applicability to Other GNN Models}
While this paper demonstrates the feasibility and efficiency of \tool using GCNs for transductive learning tasks, the proposed framework is not limited to this setting. Our design can be readily extended to support more advanced, message-passing based GNN models such as GraphSAGE, which are known for their efficient scalability and inductive generalization capabilities.

Specifically, the neighbor information padding method and the proposed secure array access protocol can be applied in e.g. GraphSAGE-based frameworks. For instance, PS-GNN~\cite{Yuan2024PS-GNN} is a privacy-preserving GNN framework designed for GraphSAGE, built on (2,2)-ASS, where the secure array access protocol is assisted by a trusted third party. Our secure array access protocol, which is designed for a four-party setting without a trusted party, can be seamlessly integrated into PS-GNN by introducing two additional helper parties. As described in \textsc{Protocol}~\ref{algo_aa}, parties $P_0$ and $P_1$ can send their shares to $P_2$ and $P_3$ for secure computation, and then receive the results to proceed with subsequent steps (see Table~\ref{table-aa-comparison} for details on the efficiency of the secure array access protocol). 

To further validate the generality of \tool, we also conduct additional experiments under \textit{inductive} settings using GraphSAGE as the backbone, and compare \tool with PS-GNN. The results (see Section~\ref{sec-inference-psgnn-exp} and TABLE~\ref{table-psgnn-inference}) demonstrate that \tool achieves improved performance while providing the same strong security guarantees without relying on a trusted third party. It is worth noting that our random neighbor information padding method and secure array access protocol are generic and can be applied to models that rely on message-passing.

\subsection{Deployment Strategies for \toolnott}
The security of \tool hinges on the non-collusion of its four parties, a critical consideration for practical deployment. To realize this assumption in practical, real-world cloud environments, several deployment strategies can be employed. 
\begin{enumerate}[leftmargin=*]
    \item \textit{Inter-Provider Deployment Strategy}. One viable deployment strategy involves distributing the four parties across independent cloud providers (e.g., GCP, Azure, AWS, Oracle). The inherent administrative and infrastructural independence of each provider naturally mitigates the risk of collusion. While this multi-cloud approach may introduce additional communication latency, it provides a strong trust model suitable for high-security applications in sectors like finance and healthcare.
    \item \textit{Single-Provider Deployment Strategy}. When operating within a single cloud provider, robust logical and physical separation can be achieved by deploying each party in a geographically distinct region or availability zone. This involves using isolated virtual machine instances complemented by stringent access control mechanisms.
    \item \textit{Federated and Inter-Organizational Deployment Strategy}. Our framework is inherently well-suited for federated or inter-organizational settings. In such scenarios—for instance, collaborations between hospitals, research institutions, or financial consortia—each participating entity naturally operates within an independent administrative domain. This alignment between organizational separation and our protocol's requirements makes it a particularly fitting model for multi-party collaborations.
\end{enumerate}

\section{Conclusion and Future Work}
In this paper, we propose a \textit{cost-effective} four-party framework, \tool, based on lightweight secret sharing techniques, designed for privacy protection and efficient training and inference of GNNs in untrusted cloud environments.
The main contributions of \tool include the introduction of a fourth-party implementation of an asynchronous computation mode for \textit{secure array access protocol}, the use of a \textit{random neighbor information padding} method to reduce the subsequent computation and communication overhead. Our experiments on the Cora, CiteSeer, and PubMed datasets validate the effectiveness of \tool, demonstrating its potential for cost-effective, privacy-preserving GNN services in the cloud.

Despite demonstrating superior performance over state-of-the-art methods for privacy-preserving GNNs, our work has several limitations. First, our secure array access protocol requires resending the entire array for each element access. This design incurs significant communication overhead, especially for large graphs. Second, our framework is currently limited to transductive learning and does not support inductive tasks, which we leave for future work. Finally, \tool assumes a semi-honest adversary model; its robustness against malicious adversaries deserves further investigation.

\section*{Acknowledgment}
This work was supported by the National Natural Science Foundation of China under Grant Nos. 62472318, 62172301, 61972241, and 62402342, the Natural Science Foundation of Shanghai under Grant No. 22ZR1427100, the Soft Science Project of Shanghai under Grant No. 25692107400, the Shanghai Pujiang Program under Grant No. 23PJ1412700, and the Shanghai Sailing Program under Grant No. 24YF2749500. Yang Shi is also with the Shanghai Engineering Research Center for Blockchain Applications and Services, and this work is also supported in part by this research center (No. 19DZ2255100) and the Shanghai Institute of Intelligent Science and Technology at Tongji University.

\bibliographystyle{IEEEtran}
\bibliography{Reference}

\begin{IEEEbiography}[{\includegraphics[width=1in,height=1.25in,clip,keepaspectratio]{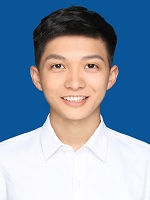}}]{Congcong Chen}
received the BE degrees in Software Engineering from Jiangxi Agricultural University, Nanchang, China, in 2018. In 2022, he earned his MSc degree in Computer Engineering from the College of Information Technology at Shanghai Ocean University, Shanghai, China. Currently, he is pursuing a Ph.D. in Software Engineering at the School of Computer Science and Technology, Tongji University, Shanghai, China. His primary research interests include GNNs, privacy-preserving machine learning, and secure multi-party computation.
\end{IEEEbiography}

\begin{IEEEbiography}[{\includegraphics[width=1in,height=1.25in,clip,keepaspectratio]{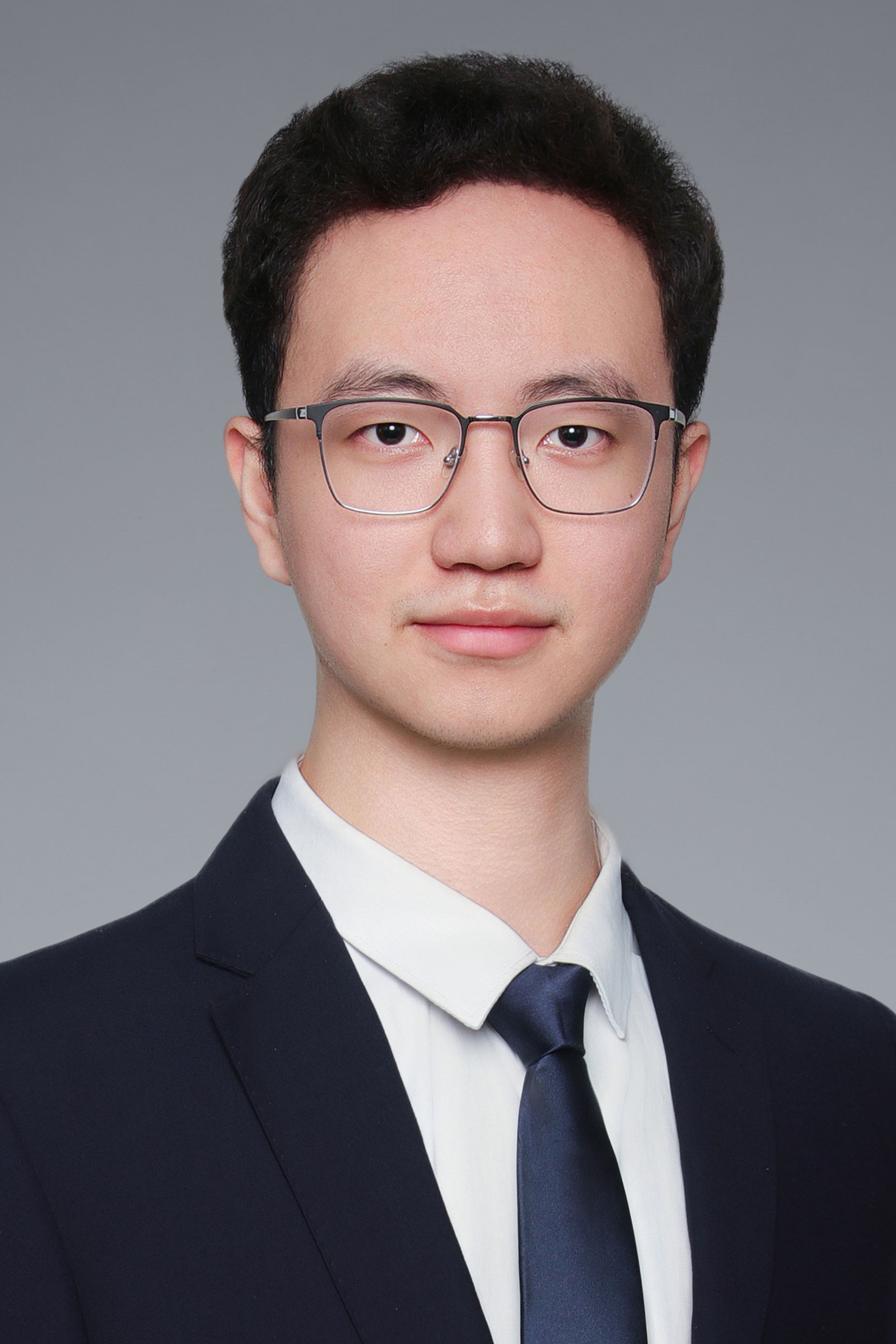}}]{Xinyu Liu}
received the BE degree in Software Engineering from Tongji University, Shanghai, China, in 2023. He is currently pursuing the Master's degree in Software Engineering at the School of Computer Science and Technology, Tongji University, Shanghai, China. His research interests include AI security, malware detection, secure multi-party computation, and large language models.
\end{IEEEbiography}

\begin{IEEEbiography}[{\includegraphics[width=1in,height=1.25in,clip,keepaspectratio]{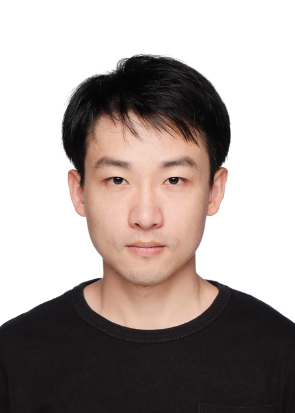}}]{Kaifeng Huang} (Member, ACM) received the Ph.D. degree in software engineering from Fudan University, China in 2022. He is currently an Assistant Professor with the School of Computer Science and Technology, Tongji University. His research interests include open-source software supply chain, software security, and software evolution.
\end{IEEEbiography}

\begin{IEEEbiography}[{\includegraphics[width=1in,height=1.25in,clip,keepaspectratio]{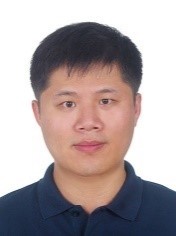}}]{Lifei Wei}
received the BSc and MSc degrees in applied mathematics from the University of Science and Technology Beijing, Beijing, China, in 2005 and 2007, respectively and the Ph.D. degree in computer science from Shanghai Jiao Tong University, Shanghai, China, in 2013. 
He is currently a Professor with College of Information Engineering, Shanghai Maritime University, Shanghai, China. He has published papers in major journals, such as IEEE Transactions on Information Forensics and Security (TIFS), IEEE Transactions on Dependable and Secure Computing (TDSC), IEEE Transactions on Services Computing, and Information Sciences. His main research interests include information security, AI security, privacy preserving, blockchain and applied cryptography. 
\end{IEEEbiography}

\begin{IEEEbiography}[{\includegraphics[width=1in,height=1.25in,clip,keepaspectratio]{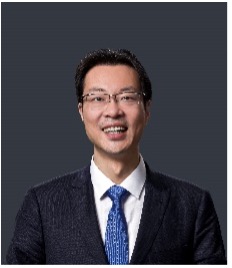}}]{Yang Shi} received his BS degree in electronic engineering from Hefei University of Technology, China, in 1999 and his PhD degree in pattern recognition and intelligent systems from Tongji University, Shanghai, China, in 2005. From 2005 to 2011, he worked for Pudong CS\&S Co. Ltd., Shanghai, China. He is now a professor of the School of Computer Science and Technology, Tongji University. His research interests include cryptography, privacy computing, AI security, blockchain technology, and software protection.
\end{IEEEbiography}

\begin{appendices}
\section{Analysis of Neighbor Information Padding Method of \toolnott}\label{sec-padding-analysis}
\textbf{Efficiency analysis.} Our \textit{random neighbor information padding} method reduces approximately $\frac{1}{2}\cdot (d_{\textit{max}}-d_{v_i})$ operations per node compared to SecGNN~\cite{wang2023secgnn}, where $d_{max}$ is the maximum degree and $d_{v_i}$ is the degree of node $v_i$. In the previous method, the larger the $d_{max}$, the greater the computational and communication overhead wasted on dummy nodes. In most real-world scenarios, however, the adjacency matrix of a graph is sparse\cite{geng2020awb,auten2020hardware,hu2020featgraph,huang2020ge,li2021gcnax}. This sparsity implies that only a small fraction of nodes have a high degree, while the vast majority have relatively low degrees. Therefore, in practice, our padding method can substantially reduce unnecessary computational and communication overhead, significantly improve the overall efficiency by roughly $\frac{1}{2}\cdot(N \cdot d_{max}+2E)$ over SecGNN, where $N$ is the number of nodes and $E$ is the number of edges in the graph $\mathcal{G}$.

\textbf{Security analysis.} Privacy-preserving GNNs aim to protect node features, edge weights and the structure of the graph from disclosure. We protected node features , edge weights, and graph structure (i.e., node neighbor IDs and node degrees) by 2-out-of-2 additive secret sharing and 2-out-of-4 replicated secret sharing techniques. To hide the structural information of the graph, we padding each node with dummy neighbor nodes, making it impossible for the adversary to learn the real graph. Unlike SecGNN~\cite{wang2023secgnn}, which pads all nodes to the maximum degree $d_{max}$, our approach uses random-sized dummy node padding, as illustrated in Fig.~\ref{fig-aa}.

For nodes with large degrees, the restricted padding space increases the likelihood of an attacker inferring their degree information. Nevertheless, as previously noted, most real-world graphs are sparse. To illustrate, TABLE \ref{table-node-distribution} presents the node degree distributions of the Cora, CiteSeer, and PubMed datasets. We can see that the number of nodes with degree greater than $\frac{1}{2} \cdot d_{max}$ is usually less than 1\% of the total number of nodes. Therefore, although it is possible to infer the degree information of nodes with large degrees, it is not possible to recover the original graph structure by knowing only the degree information of nodes with less than 1\%. Moreover, the attacker cannot accurately identify the neighbor nodes of a given node because the indexes of these neighbor nodes are encrypted using secret sharing techniques.

\begin{table}[!ht]
\centering
\caption{Node Distributions of Cora, CiteSeer and PubMed Datasets.}
\label{table-node-distribution}
\begin{threeparttable}
\begin{tabular}{cccccc}
\toprule
\multicolumn{1}{l}{\textbf{ Dataset }} & \multicolumn{1}{l}{\textbf{Nodes}} & \multicolumn{1}{l}{\textbf{Max Degree}} & \multicolumn{1}{l}{\textbf{ $d_{thre}$ }} & \multicolumn{1}{l}{\textbf{Number}} & \multicolumn{1}{l}{\textbf{Proportion}} \\ 
\midrule
\multirow{4}{*}{Cora}  & \multirow{4}{*}{2,708}  & \multirow{4}{*}{168}    & 10  & 96  & 3.55\% \\
                       &    &  & 20  & 24  & 0.89\%  \\
                       &    &   & 40    & 6    & 0.22\%      \\
                       &    &   & 60    & 4    & 0.15\%      \\
                       &    &   & 80    & 1    & 0.04\%      \\ \hline
\multirow{4}{*}{CiteSeer}   & \multirow{4}{*}{3,327}   & \multirow{4}{*}{99}   & 10  & 81    & 2.43\%   \\
                        &     &   &  20  & 13    & 0.39\%     \\
                      &     &   & 40    & 2    & 0.06\%      \\
                      &     &   & 60    & 1    & 0.03\%      \\
                     &     &    & 80    & 1    & 0.03\%      \\ \hline
\multirow{4}{*}{PubMed}    & \multirow{4}{*}{19,717}   & \multirow{4}{*}{171}   & 10   & 2328  & 11.81\%    \\
                     &    &    & 20   & 763  & 3.87\%    \\
                    &    &     & 40     & 12   & 0.64\%     \\
                  &     &     & 60   & 33     & 0.17\%     \\
                  &     &     & 80   & 8      & 0.04\%     \\ 
\bottomrule
\end{tabular}%
\begin{tablenotes} 
    \footnotesize
    \item \textbf{Note:}  $d_{thre}$ denotes the degree threshold, e.g., when $d_{thre}=10$, it means that the table will count the \textit{number} of nodes with degree greater than 10. \textit{Proportion} denotes the proportion of nodes with degree greater than $d_{thre}$ to the total number of nodes.
\end{tablenotes}
\end{threeparttable}
\end{table}

\section{Security Definitions and Functionalities}\label{functionalities}
We analyze the security of our protocol using the real-ideal paradigm \cite{canetti2000security,canetti2001universally}. This paradigm ensures that any adversary involved in a real-world interaction has a corresponding simulator in an ideal-world interaction. In other words, any information extracted by the adversary in the real-world interaction can also be extracted by the simulator in the ideal-world. The view of party $P_i$ during the execution of protocol $\Pi$ with input $\mathbf{x}$ is denoted by $\textsc{View}_i^\Pi(\mathbf{x})$. The input $\mathbf{x}$ includes the input $\textit{input}_i$ of party $P_i$, random values $\textit{random}_i$, and messages $\textit{message}_i$ received during the protocol execution. 
The definition of security is as follows:
\revise{\begin{definition}
Let $P_i, i \in \{0,1,2,3\}$ participate in protocol $\Pi$, which computes the function $f:(\{0, 1\}^{*})^{4} \rightarrow (\{0, 1\}^{*})^{4}$. 
We say that $\Pi$ securely computes function $f$ in the presence of one semi-honest corrupted party and non-colluding setting if there exists a probabilistic polynomial-time simulator $\mathcal{S}$ such that for each party $P_i$, ${\mathcal{S}(\textit{input}_i, f_i(\mathbf{x}))} \equiv {\textsc{View}_i^\Pi(\mathbf{x})}$ are computationally indistinguishable.
\end{definition}}

\revise{In this paper, we proof the security of our system using the hybrid model. In this model, parties can use sub-functionalities provided by a trusted party to execute a protocol involving real messages. The modular sequential composition theorem in \cite{canetti2001universally} demonstrates that replacing the computational sub-functionalities of a trusted party with a real secure protocol can achieve the same output distribution. When these sub-functionalities are denoted as $g$, we refer to the protocol as operating under the $g$-hybrid model. Our ideal functionalities are as follows:}

\textbf{Multiplication}. The ideal functionality of multiplication protocol $\Pi_{\mathsf{Mult}}$ is shown in Fig.~\ref{func-mult}.
\begin{figure}[h]
\begin{framed}
\centerline{\textbf{Functionality $\mathcal{F}_{\mathsf{Mult}}$}}
\raggedright
\SetKwInOut{Input}{Input}\SetKwInOut{Output}{Output}

\Input{The functionality receives inputs $\llbracket x \rrbracket$ and $\llbracket y \rrbracket$ from $P_{\{0,1,2,3\}}$.}     
\Output{Compute:}

\begin{enumerate}[label=\arabic*.]
    \item Reconstruct $x$ and $y$. 
    \item Compute $z=x\cdot y$.
    \item Generate random $(2,4)$-RSS shares of $z$.
\end{enumerate}
Send $\llbracket z \rrbracket$ back to the parties.
\end{framed}
\vspace{-0.3cm}
\caption{The multiplication ideal functionality $\mathcal{F}_{\mathsf{Mult}}$.} 
\label{func-mult}
\end{figure}

\begin{figure}[h]
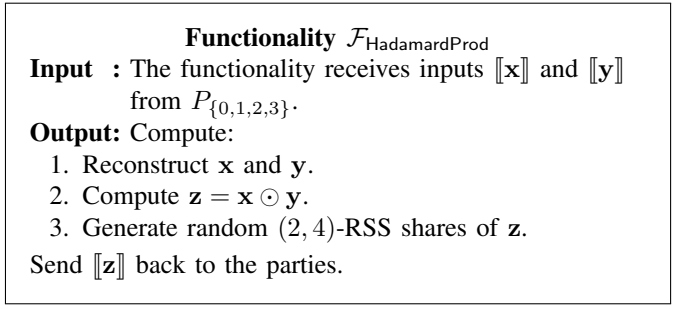

\begin{framed}
\centerline{\textbf{Functionality $\mathcal{F}_{\mathsf{HadamardProd}}$}}
\raggedright
\SetKwInOut{Input}{Input}\SetKwInOut{Output}{Output}
\Input{The functionality receives inputs $\llbracket \mathbf{x} \rrbracket$ and $\llbracket \mathbf{y} \rrbracket$ from $P_{\{0,1,2,3\}}$.} 
\Output{Compute:}

\begin{enumerate}[label=\arabic*.]
    \item Reconstruct $\mathbf{x}$ and $\mathbf{y}$. 
    \item Compute $\mathbf{z}=\mathbf{x}\odot \mathbf{y}$.
    \item Generate random $(2,4)$-RSS shares of $\mathbf{z}$.
\end{enumerate}
Send $\llbracket \mathbf{z} \rrbracket$ back to the parties.
\end{framed}
\vspace{-0.3cm}
\caption{The hadamard product ideal functionality $\mathcal{F}_{\mathsf{HadamardProd}}$.} 
\label{func-hadamard-prod}
\end{figure}

\textbf{Hadamard product}. The ideal functionality of hadamard product protocol $\Pi_{\mathsf{HadamardProd}}$ is shown in Fig.~\ref{func-hadamard-prod}. 

\begin{figure}[h]
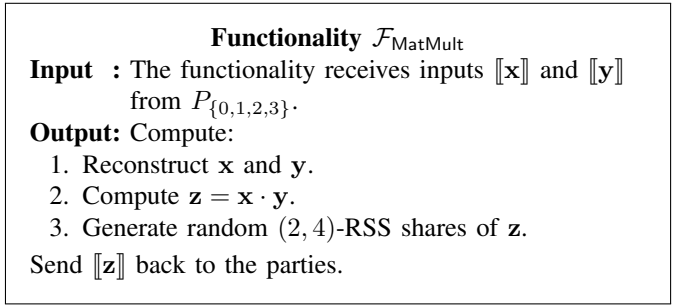

\begin{framed}
\centerline{\textbf{Functionality $\mathcal{F}_{\mathsf{MatMult}}$}}
\raggedright
\SetKwInOut{Input}{Input}\SetKwInOut{Output}{Output}
\Input{The functionality receives inputs $\llbracket \mathbf{x} \rrbracket$ and $\llbracket \mathbf{y} \rrbracket$ from $P_{\{0,1,2,3\}}$.} 
\Output{Compute:}

\begin{enumerate}[label=\arabic*.]
    \item Reconstruct $\mathbf{x}$ and $\mathbf{y}$. 
    \item Compute $\mathbf{z}=\mathbf{x}\cdot \mathbf{y}$.
    \item Generate random $(2,4)$-RSS shares of $\mathbf{z}$.
\end{enumerate}
Send $\llbracket \mathbf{z} \rrbracket$ back to the parties.
\end{framed}
\vspace{-0.3cm}
\caption{The matrix multiplication ideal functionality $\mathcal{F}_{\mathsf{MatMult}}$.} 
\label{func-matmult}
\end{figure}

\textbf{Matrix multiplication}. The ideal functionality of matrix multiplication protocol $\Pi_{\mathsf{MatMult}}$ is shown in Fig.~\ref{func-matmult}. 

\begin{figure}[h]
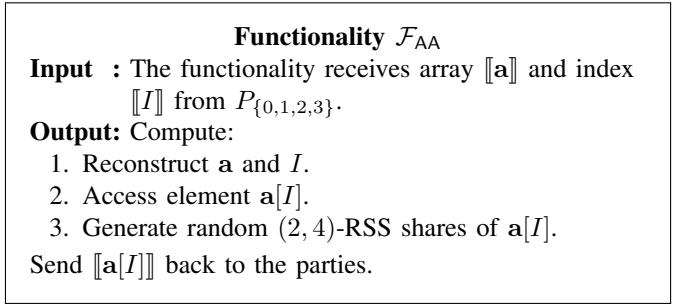

\begin{framed}
\centerline{\textbf{Functionality $\mathcal{F}_{\mathsf{AA}}$}}
\raggedright
\SetKwInOut{Input}{Input}\SetKwInOut{Output}{Output}
\Input{The functionality receives array $\llbracket \mathbf{a} \rrbracket$ and index $\llbracket I \rrbracket$ from $P_{\{0,1,2,3\}}$.} 
\Output{Compute:}

\begin{enumerate}[label=\arabic*.]
    \item Reconstruct $\mathbf{a}$ and $I$. 
    \item Access element $\mathbf{a}[I]$.
    \item Generate random $(2,4)$-RSS shares of $\mathbf{a}[I]$.
\end{enumerate}
Send $\llbracket \mathbf{a}[I] \rrbracket$ back to the parties.
\end{framed}
\vspace{-0.3cm}
\caption{The secure array access ideal functionality $\mathcal{F}_{\mathsf{AA}}$.} 
\label{func-aa}
\end{figure}

\textbf{Secure array access}. The ideal functionality of secure array access protocol $\Pi_{\mathsf{AA}}$ is shown in Fig.~\ref{func-aa}. 

\begin{figure}[h]
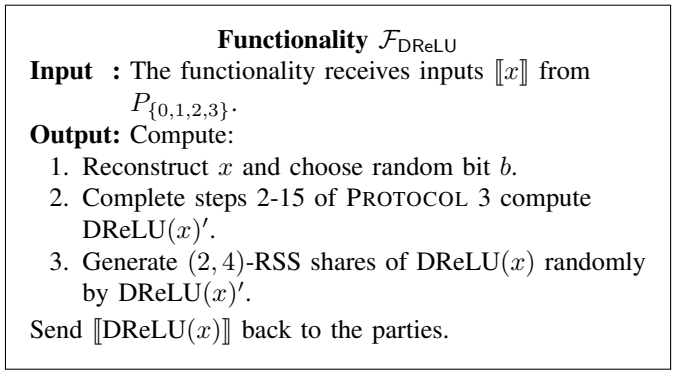

\begin{framed}
\centerline{\textbf{Functionality $\mathcal{F}_{\mathsf{DReLU}}$}}
\raggedright
\SetKwInOut{Input}{Input}\SetKwInOut{Output}{Output}
\Input{The functionality receives inputs $\llbracket x \rrbracket$ from $P_{\{0,1,2,3\}}$.} 
\Output{Compute:}

\begin{enumerate}[label=\arabic*.]
    \item Reconstruct $x$ and choose random bit $b$.
    \item Complete steps 2-15 of \textsc{Protocol} \ref{algo_drelu} compute $\text{DReLU}(x)^\prime$.
    \item Generate $(2,4)$-RSS shares of $\text{DReLU}(x)$ randomly by $\text{DReLU}(x)^\prime$.
\end{enumerate}
Send $\llbracket \text{DReLU}(x) \rrbracket$ back to the parties.
\end{framed}
\vspace{-0.3cm}
\caption{The DReLU ideal functionality $\mathcal{F}_{\mathsf{DReLU}}$.} 
\label{func-drelu}
\end{figure}

\textbf{DReLU}. The ideal functionality of DReLU protocol $\Pi_{\mathsf{DReLU}}$ is shown in Fig.~\ref{func-drelu}. 

\begin{figure}[h]
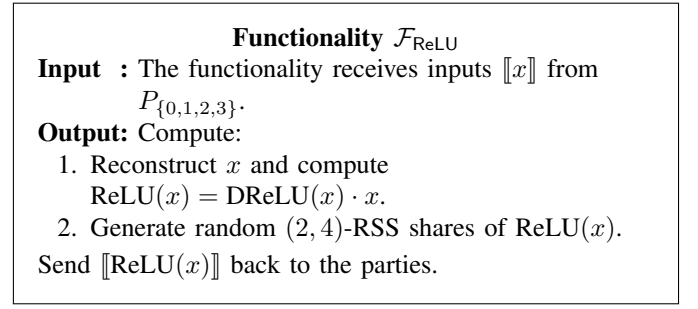

\begin{framed}
\centerline{\textbf{Functionality $\mathcal{F}_{\mathsf{ReLU}}$}}
\raggedright
\SetKwInOut{Input}{Input}\SetKwInOut{Output}{Output}
\Input{The functionality receives inputs $\llbracket x \rrbracket$ from $P_{\{0,1,2,3\}}$.} 
\Output{Compute:}

\begin{enumerate}[label=\arabic*.]
    \item Reconstruct $x$ and compute $\text{ReLU}(x)=\text{DReLU}(x)\cdot x$.
    \item Generate random $(2,4)$-RSS shares of $\text{ReLU}(x)$.
\end{enumerate}
Send $\llbracket \text{ReLU}(x) \rrbracket$ back to the parties.
\end{framed}
\vspace{-0.3cm}
\caption{The ReLU ideal functionality $\mathcal{F}_{\mathsf{ReLU}}$.} 
\label{func-relu}
\end{figure}

\textbf{ReLU}. The ideal functionality of ReLU protocol $\Pi_{\mathsf{ReLU}}$ is shown in Fig.~\ref{func-relu}. 

\begin{figure}[h]
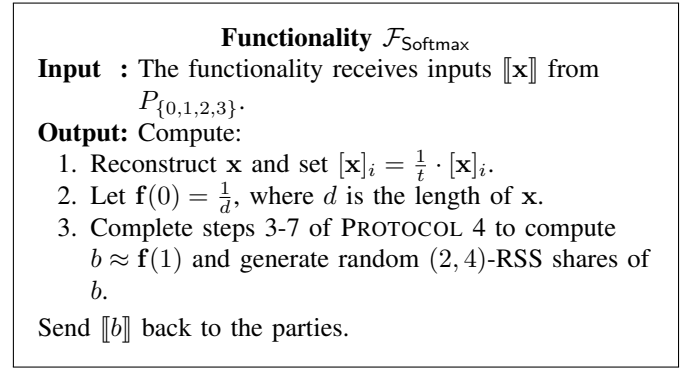

\begin{framed}
\centerline{\textbf{Functionality $\mathcal{F}_{\mathsf{Softmax}}$}}
\raggedright
\SetKwInOut{Input}{Input}\SetKwInOut{Output}{Output}
\Input{The functionality receives inputs $\llbracket \mathbf{x} \rrbracket$ from $P_{\{0,1,2,3\}}$.} 
\Output{Compute:}

\begin{enumerate}[label=\arabic*.]
    \item Reconstruct $\mathbf{x}$ and set $[\mathbf{x}]_i=\frac{1}{t} \cdot [\mathbf{x}]_i$.
    \item Let $\mathbf{f}(0)=\frac{1}{d}$, where $d$ is the length of $\mathbf{x}$.
    \item Complete steps 3-7 of \textsc{Protocol} \ref{algo_softmax} to compute $b\approx \mathbf{f}(1)$ and generate random $(2,4)$-RSS shares of $b$.
\end{enumerate}
Send $\llbracket b \rrbracket$ back to the parties.
\end{framed}
\vspace{-0.3cm}
\caption{The softmax ideal functionality $\mathcal{F}_{\mathsf{Softmax}}$.} 
\label{func-softmax}
\end{figure}

\textbf{Softmax}. The ideal functionality of softmax protocol $\Pi_{\mathsf{Softmax}}$ is shown in Fig.~\ref{func-softmax}. 

\section{Security Proofs}\label{security-proofs}
We denote $\mathcal{S}_{P_i}^{\Pi_\textsc{Protocol}}$ as the simulator for protocol $\Pi_\textsc{Protocol}$ for party $P_i$, where $i \in \{0,1,2,3\}$. To prove the security of our system, we provide simulators for $\Pi_{\mathsf{Mult}}$ (Fig.~\ref{func-mult}), $\Pi_{\mathsf{HadamardProd}}$ (Fig.~\ref{func-hadamard-prod}), $\Pi_{\mathsf{MatMult}}$ (Fig.~\ref{func-matmult}), $\Pi_{\mathsf{AA}}$ (Fig.~\ref{func-aa}), $\Pi_{\mathsf{DReLU}}$ (Fig.~\ref{func-drelu}), $\Pi_{\mathsf{ReLU}}$ (Fig.~\ref{func-relu}), and $\Pi_{\mathsf{Softmax}}$ (Fig.~\ref{func-softmax}), 
their indistinguishability is proven by Theorems \ref{theorem-mult} to \ref{theorem-softmax}. 
The proofs of Theorems \ref{theorem-mult} to \ref{theorem-softmax} are given below.

\subsection{Proof of Theorem \ref{theorem-mult}}

\textsc{Theorem 1}. \textit{Protocol $\Pi_{\mathsf{Mult}}$ securely realizes $\mathcal{F}_{\mathsf{Mult}}$ (see Fig.~\ref{func-mult}, Appendix~\ref{functionalities}) in the presence of one semi-honest corrupted party}.

\noindent \textit{Proof}. We first prove the correctness of the multiplication. Let $z_0$ and $z_1$ be the outputs of $P_0$ and $P_1$. We can derive simply (ignoring truncation here):
\begin{align}\label{eq-proof-mult}
[z]_0+[z]_1 &=([s]_0+[s]_3)+([s]_1+[s]_2) \notag \\
&=([x]_0[y]_0^{\prime}-r_{01}+[x]_3[y]_3^{\prime}+r_{23})+([x]_1[y]_1^{\prime} \notag \\ &\qquad +r_{01}+ [x]_2[y]_2^{\prime}-r_{23}) \notag \\
&=[x]_0[y]_0^{\prime}+[x]_3[y]_3^{\prime}+[x]_1[y]_1^{\prime}+[x]_2[y]_2^{\prime} \notag \\
&=[x]_0[y]_0^{\prime}+[x]_0[y]_1^{\prime}+[x]_1[y]_1^{\prime}+[x]_1[y]_0^{\prime} \notag \\
&=([x]_0+[x]_1)([y]_0^{\prime}+[y]_1^{\prime})=xy,
\end{align} 
where $[x]_3=[x]_0$, $[y]_3^{\prime}=[y]_1^{\prime}$, $[x]_2=[x]_1$, and $[y]_2^{\prime}=[y]_0^{\prime}$. The results for the other two-by-two reconstruction cases are similar to the derivation of Equation (\ref{eq-proof-mult}). For security, we consider, in turn, the simulator $P_i$, where $i \in \{0,1,2,3\}$:
\begin{itemize}[leftmargin=*]
    \item $\mathcal{S}_{P_{\{0,1,2,3\}}}^{\Pi_\textsf{Mult}}$: The simulator for $\Pi_{\mathsf{Mult}}$ can simulate transcripts in step 6 (\textsc{Protocol}~\ref{algo_mult}). The remaining steps in \textsc{Protocol}~\ref{algo_mult} are computed locally and do not require simulation. Since $[x]_i$, $[y]_i$, $[x]_i^{\prime}$, and $[y]_i^{\prime}$ are chosen uniformly at random, the outputs $[z]_i$ and $[z]_i^{\prime}$ are uniformly randomized and do not contain any information about the secrets $x$ and $y$. Therefore, the corrupted party $P_i$, where $i\in \{0,1,2,3\}$, does not receive any information.\qed
\end{itemize}

\subsection{Proof o f Theorem \ref{theorem-hadamard-prod}}

\textsc{Theorem 2}. \textit{Protocol $\Pi_{\mathsf{HadamardProd}}$ securely realizes $\mathcal{F}_{\mathsf{HadamardProd}}$ (see Fig.~\ref{func-hadamard-prod}, Appendix~\ref{functionalities}) in the presence of one semi-honest corrupted party in the ($\mathcal{F}_{\mathsf{Mult}}$)-hybrid model}.

\noindent \textit{Proof}. For $\mathcal{S}_{P_{\{0,1,2,3\}}}^{\Pi_\textsf{HadamardProd}}$, similar to the multiplication protocol, since $[\mathbf{x}]_i,[\mathbf{y}]_i,[\mathbf{x}]_i^{\prime},[\mathbf{y}]_i^{\prime}$ are uniformly chosen at random, the output $[\mathbf{z}]_i,[\mathbf{z}]_i^{\prime}$ are uniformly randomized and contain no information about the secrets $\mathbf{x}$ and $\mathbf{y}$. Similar to the protocol $\Pi_\textsf{Mult}$, the corrupted party $P_i$, where $i\in \{0,1,2,3\}$, does not retrieve any information. Protocol $\Pi_{\mathsf{HadamardProd}}$ consist of a simple combination of $\mathcal{F}_{\mathsf{Mult}}$, and the simulation is accomplished using a hybrid argument, which is therefore simulated using the corresponding simulator.\qed

\subsection{Proof of Theorem \ref{theorem-matmult}}
\textsc{Theorem 3}. \textit{Protocol $\Pi_{\mathsf{MatMult}}$ securely realizes $\mathcal{F}_{\mathsf{MatMult}}$ (see Fig.~\ref{func-matmult}, Appendix~\ref{functionalities}) in the presence of one semi-honest corrupted party in the ($\mathcal{F}_{\mathsf{Mult}}$)-hybrid model}.

\noindent \textit{Proof}. Similar to $\mathcal{F}_{\mathsf{HadamardProd}}$, protocol $\Pi_{\mathsf{MatMult}}$ can be simulated by the simulator for $\mathcal{F}_{\mathsf{Mult}}$. $[\mathbf{x}]_i,[\mathbf{y}]_i,[\mathbf{x}]_i^{\prime},$ $[\mathbf{y}]_i^{\prime}$ are uniformly chosen at random, where $i\in \{0,1,2,3\}$. Thus, the outputs $[\mathbf{z}]_i,[\mathbf{z}]_i^{\prime}$ are also uniformly random. So the output of the real protocol is indistinguishable from the view of $P_{\{0,1,2,3\}}$.
\qed

\subsection{Proof of Theorem \ref{theorem-aa}}
\textsc{Theorem 4}. \textit{Protocol $\Pi_{\mathsf{AA}}$ securely realizes $\mathcal{F}_{\mathsf{AA}}$ (see Fig.~\ref{func-aa}, Appendix~\ref{functionalities}) in the presence of one semi-honest corrupted party in the ($\mathcal{F}_{\mathsf{SC}}$)-hybrid model}.

\noindent \textit{Proof}. We consider the simulators for $P_i$, where $i\in \{0,1,2,3\}$, respectively:
\begin{itemize}[leftmargin=*]
    \item $\mathcal{S}_{P_{\{2,3\}}}^{\Pi_\textsf{AA}}$: In Equation (\ref{aa-element-rotation}), $[\mathbf{a}]_i$ and $[I]_i$ are rotated by $r_{01}^{a}$, and the elements of $[\mathbf{a}]_i$ are masked by $\mathbf{r}_{01}^{b}$, where $i \in \{0,1\}$. Since $P_2$ and $P_3$ do not have any information about $r_{01}^{a}$ and $\mathbf{r}_{01}^{b}$ and are chosen uniformly at random by $P_0$ and $P_1$, only the masked $\mathbf{a}^{\prime\prime}$ and index $h$ are obtained by $P_2$ and $P_3$. Thus, $\mathbf{a}^{\prime\prime}$ and index $h$ are uniformly randomized in the views of $P_2$ and $P_3$.
    \item $\mathcal{S}_{P_{\{0,1\}}}^{\Pi_\textsf{AA}}$: We can see that $P_0$ received $[\mathbf{a}[h]]_1^{\prime\prime} - r_{23}^{a}$ and $[\mathbf{a}[h]]_0^{\prime\prime} + r_{23}^{a} - r_{23}^{b}$, and $P_1$ received $[\mathbf{a}[h]]_0^{\prime\prime} + r_{23}^{a}$ and $[\mathbf{a}[h]]_1^{\prime\prime} - r_{23}^{a} + r_{23}^{b}$. $r_{23}^{a}$ and $r_{23}^{b}$ are chosen randomly and uniformly by $P_2$ and $P_3$ and are not known by $P_0$ and $P_1$. Therefore, $[\mathbf{a}[I]]_i$ and $[\mathbf{a}[I]]_i^\prime$, where $i\in \{0,1\}$, are uniformly randomized in the views of $P_0$ and $P_1$.
\end{itemize}
Hence, the final output ($[\mathbf{a}[I]]_i,[\mathbf{a}[I]]_i^\prime$), where $i\in \{0,1,2,3\}$, are also randomized and contain no information about $\mathbf{a}$. Step~1 in \textsc{Protocol} \ref{algo_aa} is simulated by the simulator of $\mathcal{F}_{\mathsf{SC}}$. The simulator for $\Pi_{\mathsf{AA}}$ can simulate transcripts in steps 3, 4, 6, and 7 (\textsc{Protocol} \ref{algo_aa}). Other steps are computed locally, so there is no need for simulation.
\qed

\subsection{Proof of Theorem \ref{theorem-drelu}}
\textsc{Theorem 5}. \textit{Protocol $\Pi_{\mathsf{DReLU}}$ securely realizes $\mathcal{F}_{\mathsf{DReLU}}$ (see Fig.~\ref{func-drelu}, Appendix~\ref{functionalities}) in the presence of one semi-honest corrupted party in the ($\mathcal{F}_{\mathsf{MS}}$)-hybrid model}.

\noindent \textit{Proof}. Similar to the previous proof procedure, steps 12, 14, and 18 of \textsc{Protocol}~\ref{algo_drelu} can be simulated using the simulator for $\Pi_{\mathsf{DReLU}}$. Other steps in \textsc{Protocol}~\ref{algo_drelu} do not require simulation. The simulators for $P_i$, where $i\in \{0,1,2,3\}$, are as follows:
\begin{itemize}[leftmargin=*]
    \item $\mathcal{S}_{P_{\{0,1\}}}^{\Pi_\textsf{DReLU}}$: The $[\text{DReLU}(x)^{\prime\prime}]_i$, where $i\in \{0,1\}$, received by $P_0$ and $P_1$ is masked by the randomly chosen $r$ of $P_2$ or $P_3$. Thus, for the views of $P_0$ and $P_1$, the real protocol outputs of $[\text{DReLU}(x)]_j$ and $[\text{DReLU}(x)]_j^\prime$, where $j\in \{0,1\}$, have identical distributions to the simulator's outputs.
    \item $\mathcal{S}_{P_{\{2, 3\}}}^{\Pi_\textsf{DReLU}}$: $P_2$ or $P_3$ first receives $[\mathbf{w}]_0$ and $[\mathbf{w}]_1$ from $P_0$ and $P_1$, while $b$ and $\mathbf{r}_{01}$ are chosen uniformly at random by $P_0$ and $P_1$. Therefore, they do not contain any information about the secret $x$. Finally, $P_2$ and $P_3$ receive $[\text{DReLU}(x)]_j$ and $[\text{DReLU}(x)]_j^\prime$, where $j\in \{2,3\}$, which is masked by $r_{01}^a$ and $r_{01}^b$. Since $r_{01}^a$ and $r_{01}^b$ are also chosen uniformly at random by $P_0$ and $P_1$, the distribution of the output of the real protocol $\Pi_{\mathsf{DReLU}}$ and the simulator generated $[\text{DReLU}(x)]_j$ and $[\text{DReLU}(x)]_j^\prime$, where $j\in \{2,3\}$, are identical from the view of $P_2$ or $P_3$. \qed
\end{itemize}

\subsection{Proof of Theorem \ref{theorem-relu}}
\textsc{Theorem 6}. \textit{Protocol $\Pi_{\mathsf{ReLU}}$ securely realizes $\mathcal{F}_{\mathsf{ReLU}}$ (see Fig.~\ref{func-relu}, Appendix~\ref{functionalities}) in the presence of one semi-honest corrupted party in the ($\mathcal{F}_{\mathsf{Mult}}$, $\mathcal{F}_{\mathsf{DReLU}}$)-hybrid model}.

\noindent \textit{Proof}. The protocol $\Pi_{\mathsf{ReLU}}$ consists of $\mathcal{F}_{\mathsf{Mult}}$, $\mathcal{F}_{\mathsf{DReLU}}$, and local computation. The simulators for $\mathcal{F}_{\mathsf{Mult}}$ and $\mathcal{F}_{\mathsf{DReLU}}$ compose the simulation $\Pi_{\mathsf{ReLU}}$. 
Thus, we can easily proof that the output of the real protocol and the view of $P_{\{0,1,2,3\}}$ are identical. \qed

\subsection{Proof of Theorem \ref{theorem-softmax}}
\textsc{Theorem 7}. \textit{Protocol $\Pi_{\mathsf{Softmax}}$ securely realizes $\mathcal{F}_{\mathsf{Softmax}}$ (see Fig.~\ref{func-softmax}, Appendix~\ref{functionalities}) in the presence of one semi-honest corrupted party in the ($\mathcal{F}_{\mathsf{Mult}}$, $\mathcal{F}_{\mathsf{HadamardProd}}$)-hybrid model}.

\noindent \textit{Proof}. The protocol $\Pi_{\mathsf{Softmax}}$ consists of $\mathcal{F}_{\mathsf{Mult}}$, $\mathcal{F}_{\mathsf{HadamardProd}}$, and local computation. The simulation is performed by sequentially composing the simulators for $\mathcal{F}_{\mathsf{Mult}}$ and $\mathcal{F}_{\mathsf{HadamardProd}}$. Step 4 of \textsc{Protocol} \ref{algo_softmax} can be simulated using the simulator for $\mathcal{F}_{\mathsf{Mult}}$, while step 5 can be simulated using the simulator for $\mathcal{F}_{\mathsf{HadamardProd}}$. The remaining steps involve local computation and do not require simulation. Similarly to $\Pi_{\mathsf{ReLU}}$, we can prove that the output of the real protocol is consistent with the view of $P_{\{0,1,2,3\}}$.\qed

\section{Theoretical Complexity}\label{sec-appendix-theoretical}
We estimated the theoretical complexity of the communication overhead of the protocols in \tool, as shown in TABLE \ref{table-theoretical-complexity}. 
In addition to $\Pi_\textsf{DReLU}$, $\Pi_\textsf{ReLU}$ incurs additional communication overhead for multiplication computation. The $\Pi_\textsf{Softmax}$ and $\Pi_\textsf{InvSqrt}$ mainly come from the communication overhead of multiplication computation.
\begin{table}[!ht]
\centering
\resizebox{\linewidth}{!}{%
\begin{threeparttable}
\caption{Theoretical Complexity of Protocols in \toolnott.}
\label{table-theoretical-complexity}
\begin{tabular}{cccc}
\toprule
\textbf{Protocol} & \textbf{Dependence} & \textbf{Rounds} & \textbf{Comm. Overhead (bits)$^\ast$} \\ 
\midrule
\textsf{Mult}    & $-$                      & $1$          & $4\ell$ \\
\textsf{HadamardProd} & $n$                      & $1$          & $4n\ell$ \\
\textsf{MatMult} & $(x\times y)(y\times z)$ & $1$          & $4xz\ell$ \\
\textsf{AA}      & $n$                      & $1$          & $(6+2n)\ell$ \\
\textsf{DReLU}   & $n$                      & $2$          & $n(\ell_x+1)(\ell_x+1)+6n\ell$ \\
\textsf{ReLU}    & $n$                      & $3$          & $n(\ell_x+1)(\ell_x+1)+10n\ell$ \\
\textsf{Softmax} & $n,t$                    & $3t$         & $12nt\ell$ \\
\textsf{InvSqrt} & $n,r$ & $3r+\text{log}_2\ell$  & \begin{tabular}[c]{@{}c@{}}$n(\ell_x+1)(\ell_x+1)\cdot\text{log}_2\ell$\\ $+7n\ell\cdot\text{log}_2\ell+12nr\ell$\end{tabular}\\
\bottomrule
\end{tabular}%
\begin{tablenotes} 
    \footnotesize
    \item \textbf{Notations:} $\ell$ - size of ring in bits. $n$ - size of vectors. $x,y,z$ - size of the matrix. $\ell_x$ - key bits of the $x$, which in this paper is 13 bits. $t$ and $r$ - hyperparameters for the softmax and inverse square root loop iteration. 
    \item \textbf{$^\ast$}: \textit{Comm. Overhead } indicates communication overhead.
\end{tablenotes}
\end{threeparttable}
}
\end{table}

\section{Implementation Optimizations}\label{sec-appendix-impl}

\textbf{Matrix Optimization}. We used the Eigen library (version 3.3.4) for fast matrix-vector operations, which significantly improved CPU performance.

\textbf{DReLU Optimization}. To reduce computational and communication overhead in the DReLU protocol, we also adopt the key-bits technique from Bicoptor 2.0~\cite{zhou2023bicoptor2}, choosing a 13-bit configuration (11 integer, 2 fractional bits).

\textbf{Precomputation of Neighbor States}. During both training and inference, the initial neighbor states for each node are static. We precompute these states in \tool and SecGNN~\cite{wang2023secgnn} and store them in files before secure computation. This greatly reduces online computation and communication without affecting GNN accuracy.

\section{Hyperparameter Analysis of \textsf{Softmax} and \textsf{InvSqrt} Protocols}

To assess the trade-off between accuracy and performance, we analyze the impact of loop iteration parameters $t$ (for \textsf{Softmax}) and $r$ (for \textsf{InvSqrt}). TABLE~\ref{table-softmax-analysis} and TABLE~\ref{table-invsqrt-analysis} present the Root Mean Square Error (RMSE), computational time, and communication overhead across different values of $t$ and $r$. 10, 100 and 500 indicate the size of the value used for testing.

\subsubsection{\textsf{Softmax} ($t$-iterations)}
As shown in TABLE~\ref{table-softmax-analysis}, increasing the number of loop iterations $t$ reduces the RMSE significantly. For instance, RMSE drops from $4.17 \times 10^{-3}$ at $t=1$ to $5.63 \times 10^{-4}$ at $t=8$, with a relatively moderate increase in both computation time and communication overhead. After $t=8$, further improvements in RMSE become marginal (e.g., $7.86 \times 10^{-4}$ at $t=16$ and $1.46 \times 10^{-3}$ at $t=32$) while time and communication overhead rise sharply. Therefore, setting $t=8$ achieves a good trade-off between accuracy and efficiency.

\subsubsection{\textsf{InvSqrt} ($r$-iterations)}
Similarly, TABLE~\ref{table-invsqrt-analysis} indicates that $r=4$ yields an RMSE of $1.82 \times 10^{-3}$, improving significantly over $r=1$ or $r=2$, and closely approximating the RMSE of larger $r$ values (e.g., $1.35 \times 10^{-3}$ at $r=16$). Meanwhile, time and communication overhead remain low at $r=4$, especially compared to $r=16$ or $r=32$. Thus, $r=4$ provides sufficient numerical accuracy without incurring excessive resource consumption.

Therefore, based on the analysis in TABLE~\ref{table-softmax-analysis} and TABLE~\ref{table-invsqrt-analysis}, we set the \textsf{Softmax} loop iteration parameter to $t=8$ and the \textsf{InvSqrt} loop parameter to $r=4$ throughout our main experiments to ensure that the secure protocols remain practically efficient while maintaining high numerical fidelity.

\begin{table*}[!ht]
\color{revise-color}
\centering
\caption{Hyperparameter Analysis of \textsf{Softmax} Protocol.}
\label{table-softmax-analysis}
\begin{tabular}{cccc|ccc|ccc}
\toprule
\multirow{2}{*}{\textbf{Loop} ($t$)} & \multicolumn{3}{c|}{\textbf{10 values}} & \multicolumn{3}{c|}{\textbf{100 values}} & \multicolumn{3}{c}{\textbf{500 values}} \\
\cmidrule(r){2-4} \cmidrule(r){5-7} \cmidrule(r){8-10}
 & \textbf{RMSE} & \textbf{Time (s)} & \textbf{Comm. (MB)} & \textbf{RMSE} & \textbf{Time (s)} & \textbf{Comm. (MB)} & \textbf{RMSE} & \textbf{Time (s)} & \textbf{Comm. (MB)} \\
\midrule
1  & $4.17\times 10^{-3}$ &  0.2863 & 0.0065 & $3.60\times 10^{-4}$ & 0.5587 & 0.0710 & $9.30\times 10^{-5}$ & 0.7195 & 0.3580 \\
2  & $2.23\times 10^{-3}$ & 0.3683 & 0.0104 & $1.71\times 10^{-4}$ & 0.6339 & 0.1094 & $8.10\times 10^{-5}$ & 0.8117 & 0.5500 \\
4  & $1.10\times 10^{-3}$ & 0.4920 & 0.0180 & $9.40\times 10^{-5}$ & 0.7556 & 0.1862 & $5.90\times 10^{-5}$ & 0.9304 & 0.9340 \\
8  & $\mathbf{5.63\times 10^{-4}}$ & 0.7387 & 0.0334 & $\mathbf{6.60\times 10^{-5}}$ & 1.0066 & 0.3398 & $4.30\times 10^{-5}$ & 1.1684 & 1.7020 \\
16 & $7.86\times 10^{-4}$ & 1.2422 & 0.0641 & $8.30\times 10^{-5}$ & 1.4947 & 0.6470 & $3.80\times 10^{-5}$ & 1.6587 & 3.2380 \\
32 & $1.46\times 10^{-3}$ & 2.2210 & 0.1256 & $1.15\times 10^{-4}$ & 2.5093 & 1.2614 & $\mathbf{3.40\times 10^{-5}}$ & 2.6529 & 6.3100 \\
\bottomrule
\end{tabular}
\end{table*}

\begin{table*}[!ht]
\color{revise-color}
\centering
\caption{Hyperparameter Analysis of \textsf{InvSqrt} Protocol.}
\label{table-invsqrt-analysis}
\begin{tabular}{cccc|ccc|ccc}
\toprule
\multirow{2}{*}{\textbf{Loop} ($r$)} & \multicolumn{3}{c|}{\textbf{10 values}} & \multicolumn{3}{c|}{\textbf{100 values}} & \multicolumn{3}{c}{\textbf{500 values}} \\
\cmidrule(r){2-4} \cmidrule(r){5-7} \cmidrule(r){8-10}
 & \textbf{RMSE} & \textbf{Time (s)} & \textbf{Comm. (MB)} & \textbf{RMSE} & \textbf{Time (s)} & \textbf{Comm. (MB)} & \textbf{RMSE} & \textbf{Time (s)} & \textbf{Comm. (MB)} \\
\midrule
1  & $6.53\times 10^{-2}$ & 0.4751 & 0.0202 & $6.09\times 10^{-2}$ & 0.4707 & 0.2016 & $6.68\times 10^{-2}$ & 0.4730 & 1.0080 \\
2  & $1.53\times 10^{-2}$ & 0.5117 & 0.0240 & $1.65\times 10^{-2}$ & 0.5120 & 0.2400 & $2.71\times 10^{-2}$ & 0.5209 & 1.2000 \\
4  & $1.82\times 10^{-3}$ & 0.6398 & 0.0317 & $\mathbf{8.49\times 10^{-4}}$ & 0.6451 & 0.3168 & $\mathbf{9.65\times 10^{-4}}$ & 0.6373 & 1.5840 \\
8  & $\mathbf{1.28\times 10^{-3}}$ & 0.8958 & 0.0470 & $1.34\times 10^{-3}$ & 0.8809 & 0.4704 & $1.23\times 10^{-3}$ & 0.8908 & 2.3520 \\
16 & $1.35\times 10^{-3}$ & 1.3728 & 0.0778 & $1.84\times 10^{-3}$ & 1.3877 & 0.7776 & $1.81\times 10^{-3}$ & 1.3802 & 3.8880 \\
32 & $1.51\times 10^{-3}$ & 2.3808 & 0.1392 & $2.51\times 10^{-3}$ & 2.3670 & 1.3920 & $2.70\times 10^{-3}$ & 2.3854 & 6.9600 \\
\bottomrule
\end{tabular}
\end{table*}

\end{appendices}
\end{document}